\documentclass[aps,prd,twocolumn,superscriptaddress,nofootinbib,floatfix]{revtex4-2}

\usepackage[T1]{fontenc}
\usepackage[utf8]{inputenc}
\usepackage{lmodern}
\usepackage[reqno]{amsmath}
\usepackage{amssymb,bm,mathtools}
\usepackage{graphicx}
\newcommand{\safeincludegraphics}[2][]{%
  \IfFileExists{#2}%
    {\includegraphics[#1]{#2}}%
    {\setlength{\fboxsep}{10pt}%
     \fbox{\parbox[c][0.30\textheight][c]{0.88\linewidth}{%
       \centering
       \normalsize Figure placeholder\\[4pt]
       \texttt{\detokenize{#2}}\\[4pt]
       \small (file not found; placeholder shown so the document compiles)%
     }}}%
}
\usepackage{tikz}
\usetikzlibrary{arrows.meta}
\usetikzlibrary{positioning,calc}
\usetikzlibrary{shapes.geometric}
\definecolor{scanblue}{rgb}{0.26,0.58,0.76}
\definecolor{loblue}{rgb}{0.13,0.40,0.67}
\definecolor{hored}{rgb}{0.70,0.09,0.17}
\definecolor{starfill}{rgb}{1.0,0.93,0.0}
\newsavebox{\figscanbox}
\newsavebox{\fighistbox}
\newsavebox{\figanglebox}
\usepackage{booktabs}
\usepackage{longtable}
\usepackage{hyperref}
\hypersetup{colorlinks=true,linkcolor=blue,citecolor=blue,urlcolor=blue}
\allowdisplaybreaks

\newcommand{\Bpar}{B}
\newcommand{\e}{\epsilon}
\newcommand{\UPMNS}{U_{\mathrm{PMNS}}}
\newcommand{\Ue}{U_e}
\newcommand{\Unu}{U_\nu}
\newcommand{\Ye}{Y_e}

\newcommand{\diag}{\mathrm{diag}}
\newcommand{\JCP}{J_{\mathrm{CP}}}
\newcommand{\mnu}{m_\nu}

\begin{document}

\title{\texorpdfstring{Lepton Mixing from a Lattice Flavon Model:\\A Two-Branch Octant--$\delta$ Prediction}{Lepton Mixing from a Lattice Flavon Model: A Two-Branch Octant--delta Prediction}}

\author{Vernon Barger}
\affiliation{Department of Physics, University of Wisconsin--Madison, Madison, Wisconsin 53706, USA}

\date{November 6, 2026}

\begin{abstract}
We extend the single-flavon $B$-lattice Froggatt--Nielsen (FN) framework, previously successful for quark masses and Cabibbo-Kobayashi-Maskawa (CKM) mixing, to the lepton sector. The same $B$-lattice power structure ($\epsilon\equiv 1/B\simeq 0.19$) generates charged-lepton mass hierarchies and a normal-ordered neutrino spectrum; large neutrino mixing angles require an additional mu--tau reflection symmetry of the neutrino mass matrix, which fixes the atmospheric angle to $45^\circ$ and the neutrino-sector Dirac phase to $\delta_\nu=270^\circ$ exactly while leaving the reactor angle, generated at $\mathcal{O}(\epsilon)$, free. The Pontecorvo-Maki-Nakagawa-Sakata (PMNS) matrix factorizes as $U_{\rm PMNS}=U_e^\dagger U_\nu$, with near-tribimaximal $U_\nu$ corrected by small charged-lepton rotations whose $12$ and $23$ phases are aligned. A single interference relation expresses the observed Dirac phase $\delta$ as the maximal neutrino phase shifted by a calculable charged-lepton correction, and correlates the sign of that shift with the atmospheric octant; this produces a two-branch prediction in the $(\theta_{23},\delta)$ plane, two lobes of one continuous distribution: a lower-octant branch with $\theta_{23}\approx 44^\circ$, $\delta\approx 260^\circ$, and an upper-octant branch with $\theta_{23}\approx 46^\circ$, $\delta\approx 275^\circ$. The structure has a geometric form as a $\nu_3$-column normalization triangle whose base angle is $2\theta_{23}$, with maximal mixing the $90^\circ$ limit. The lower octant contains the strict single-source point of the flavon phase. Both branches place $\delta$ within about $15^\circ$ of $270^\circ$, i.e.\ near-maximal leptonic CP violation, favored by T2K; a Dirac phase in the first or second quadrant, or more than about $40^\circ$ from maximal, would exclude the framework. The Jarlskog invariant $J_{\rm CP}\simeq -0.033$ is nearly branch-independent; only precision measurements of the atmospheric octant and Dirac phase at DUNE, Hyper-Kamiokande, IceCube, and JUNO can distinguish the two branches.
\end{abstract}

\maketitle

\section{Introduction}
\begin{figure}[htbp]
\centering
\begin{tikzpicture}[
    font=\small,
    box/.style={draw, rounded corners=2pt, align=center, inner sep=6pt},
    arr/.style={-Latex, line width=0.7pt},
    node distance=7mm
]
\node[box] (textures) {Single-$B$ lattice textures\\[-1pt]
$\Ye\sim \epsilon^{p^e}$,\quad $\mnu\sim \epsilon^{p^\nu}$\\[-1pt]
($\epsilon=1/B$,\; $B$-lattice power counting)};

\node[box, below=of textures] (diag) {Diagonalize\\[-1pt]
$\Ue^\dagger \Ye \Ue=\Ye^{\rm diag}$\\[-1pt]
$\Unu^T \mnu \Unu=\mnu^{\rm diag}$};

\node[box, below=of diag] (pmns) {$\UPMNS=\Ue^\dagger \Unu$\\[-1pt]
(neutrino-dominant $\Unu$ + small $\Ue$)};

\node[box, below=of pmns] (interf) {Interference in $U_{e3}$\\[-1pt]
$U_{e3}\simeq s_{13}^\nu e^{-i\delta_\nu}-\theta_{12}^e s_{23}^\nu e^{i\phi_{12}^e}$};

\node[box, below=of interf] (pred) {Prediction: octant--$\delta$ correlation\\[-1pt]
two branches in $(\theta_{23},\delta)$\\[-1pt]
selected by the same phase controlling $U_{e3}$};

\draw[arr] (textures) -- (diag);
\draw[arr] (diag) -- (pmns);
\draw[arr] (pmns) -- (interf);
\draw[arr] (interf) -- (pred);
\end{tikzpicture}
\caption{Schematic logic of the single-$B$ lattice-flavon lepton analysis.
Neutrino-dominant mixing corrected by $B$-lattice charged-lepton rotations
produces an interference-controlled reactor element $U_{e3}$, implying a
predictive two-branch octant--$\delta$ structure.}
\label{fig:schematic}
\end{figure}
The Froggatt--Nielsen mechanism~\cite{Froggatt:1978nt,Leurer1992,Leurer1993} offers a framework for addressing fermion mass hierarchies through a single expansion parameter.
In this work we develop the lepton sector of the lattice-flavon (``ninths'') implementation~\cite{Barger2025bfn,Barger2025bfnb,LatticeFlavonQuarkMixing}, in which fermion mass matrices are organized by powers of 
\(\epsilon = 1/B\), with \(B = 75/14\). 
The same single-\(B\) structure that successfully accounts for quark masses and CKM mixing~\cite{Cabibbo,KobayashiMaskawa} is applied here to charged leptons and neutrinos, supplemented in the neutrino sector by a mu--tau reflection symmetry that produces the maximal atmospheric angle, the maximal neutrino-sector phase, and, with the lattice hierarchy, the large solar angle.

The observed pattern of lepton mixing (large solar and atmospheric angles together with a comparatively small but nonzero reactor angle and a sizable Dirac CP phase~\cite{NuFIT60}) provides a rich testing ground for this framework.

Lepton mixing factorizes as~\cite{Pontecorvo,MNS}
\begin{equation}
U_{\rm PMNS} = U_e^\dagger U_\nu,
\end{equation}
so that the observed angles and phases arise from controlled interference between neutrino-dominant mixing and small charged-lepton rotations fixed by $B$-lattice power counting. 
This structure leads to a predictive linkage: the same interference that generates \(U_{e3}\) correlates the atmospheric-octant selection with the displacement of the Dirac phase \(\delta\) (Fig.~\ref{fig:schematic}).

We derive analytic diagonalization formulas, perform representative coefficient scans consistent with NuFIT~6.0~\cite{NuFIT60} normal ordering, and present a compact analytic relation, $\delta\simeq\delta_\nu-\arg(1-r\,e^{-i\Phi_e})$ [Eq.~\eqref{eq:delta-theorem}], that expresses the observed Dirac phase as the neutrino phase $\delta_\nu=270^\circ$, fixed by the reflection symmetry, shifted by a charged-lepton interference, and encodes the two-branch octant--\(\delta\) structure. 
Both branches predict near-maximal leptonic CP violation, with the Dirac phase within about $15^\circ$ of $270^\circ$ ($\delta\approx260^\circ$ in the lower octant, $275^\circ$ in the upper), the regime favored by T2K~\cite{T2K}.
The resulting correlation provides a sharp and experimentally testable signature of the single-\(B\) lattice-flavon organizing principle at DUNE~\cite{DUNE}, Hyper-Kamiokande~\cite{HyperK}, IceCube~\cite{IceCube}, and JUNO~\cite{JUNO}.
\section{Flavon Lattice and Messenger Framework}

The flavor structure is generated by a single Froggatt–Nielsen flavon field~\cite{Froggatt:1978nt,Leurer1992,Leurer1993} $\Phi$ with
$\langle \Phi \rangle / \Lambda = 1/B \equiv \epsilon$.
Effective Yukawa operators arise from integrating out heavy vectorlike messenger chains,
producing suppression factors of the form
\begin{equation}
Y_{ij} \sim c_{ij}\, \e^{\,p_{ij}}\, e^{i\phi_{ij}},
\end{equation}
where $p_{ij}$ are rational lattice exponents determined by messenger insertions,
$c_{ij}=\mathcal{O}(1)$ coefficients, and $\phi_{ij}$ are physical phases.

\subsection*{Exponent Structure}

In the charged-lepton sector, the additive charge rule
\begin{equation}
p^e_{ij} = p^e_{L_i} + p^e_{R_j}
\end{equation}
generates hierarchical masses controlled by the lattice exponents,
with the specific charge assignments detailed in Appendix~\ref{app:charges-scan} [Eq.~\eqref{eq:lepton-charges}].

The neutrino mass matrix (for normal ordering) follows
\begin{equation}
m_1 : m_2 : m_3 \sim B^{-2} : B^{-1} : 1.
\end{equation}

\subsection*{Phase Structure}

Physical CP violation arises from relative phases between entries.
After field redefinitions removing unphysical phases,
the remaining Dirac phase appears through interference of left-handed rotations:
\begin{equation}
U_{\rm PMNS} = U_e^\dagger U_\nu .
\label{eq:PMNS-factorize}
\end{equation}

In the Fritzsch--Xing (FX) convention~\cite{Fritzsch} adopted here for the lepton sector, the Dirac phase is associated with the
$U_{e3}$ element (in direct analogy with the CKM choice where the FX phase resides in $V_{us}$).
Writing the FX parameterization schematically as
\begin{equation}
U_{\rm PMNS}^{\rm (FX)} =
\begin{aligned}[t]
& R_{12}^T(\theta_\ell)\,P_\phi\,R_{23}(\theta)\,
R_{12}(\theta_\nu),
\\
& P_\phi\equiv \diag(1,\,e^{i\phi_{\rm FX}},\,1).
\end{aligned}
\end{equation}
one sees explicitly that the leptonic Jarlskog invariant~\cite{Jarlskog} takes the form
\begin{equation}
J_{\rm CP} = s_\ell c_\ell\, s_\nu c_\nu\, s^2 c\, \sin\phi_{\rm FX},
\end{equation}

\paragraph*{Relation to the Particle Data Group (PDG) phase~\cite{PDG2024}.}
Because the Jarlskog invariant is basis independent, one may equate the FX and PDG forms,
\begin{equation}
J_{\rm CP}^{\rm (FX)} = J_{\rm CP}^{\rm (PDG)},
\end{equation}
where in the PDG convention
\begin{equation}
J_{\rm CP}^{\rm (PDG)} =
s_{12} c_{12}\, s_{23} c_{23}\, s_{13} c_{13}^2\, \sin\delta_{\rm PDG}.
\end{equation}
Hence
\begin{equation}
s_\ell c_\ell\, s_\nu c_\nu\, s^2 c\, \sin\phi_{\rm FX}
=
s_{12} c_{12}\, s_{23} c_{23}\, s_{13} c_{13}^2\, \sin\delta_{\rm PDG}.
\end{equation}
The invariant physical quantity is $J_{\rm CP}$; the parameter phases $\phi_{\rm FX}$ and
$\delta_{\rm PDG}$ coincide only when the angle definitions match exactly.

\subsection*{Diagonalization}

Given a Yukawa matrix $Y$,
biunitary diagonalization yields
\begin{equation}
U_L^\dagger\, Y\, U_R = Y^{\rm diag}.
\end{equation}

\noindent
\emph{Remark.} While only the left-handed rotation matrices $U_L$ enter directly into the observable
PMNS matrix, the right-handed matrices $U_R$ and their associated angles contribute to the
diagonalization of the Yukawa operator and therefore enter the physical charged-lepton mass
eigenvalues.  In hierarchical $B$-lattice textures the scaling of $m_\ell$ depends on the combined
left- and right-handed exponents, even though $U_R$ drops out of the low-energy mixing matrix.
\paragraph*{Illustrative $2\times2$ example.}
To see how the additive charge rule translates into mixing angles, consider a
$2\times2$ charged-lepton sub-block with lattice exponents,
\begin{equation}
p \;=\;
\begin{pmatrix}
p_{L_1}+p_{R_1} & p_{L_1}+p_{R_2} \\
p_{L_2}+p_{R_1} & p_{L_2}+p_{R_2}
\end{pmatrix},
\end{equation}
\begin{equation}
Y \sim
\begin{pmatrix}
B^{-p_{11}/9} & B^{-p_{12}/9}\\
B^{-p_{21}/9} & B^{-p_{22}/9}
\end{pmatrix}.
\end{equation}
If $p_{22}<p_{12}$ and $p_{22}<p_{21}$, the dominant entry is $Y_{22}$ and the left- and right-handed
mixing angles obtained by biunitary diagonalization scale as
\begin{equation}
\begin{aligned}
\theta_{12}^{L}&\sim \frac{Y_{12}}{Y_{22}}\sim B^{-(p_{12}-p_{22})/9},\\
\theta_{12}^{R}&\sim \frac{Y_{21}}{Y_{22}}\sim B^{-(p_{21}-p_{22})/9}.
\end{aligned}
\end{equation}
Thus the same additive rule that fixes the diagonal suppressions (and hence mass ratios) automatically controls
the rotation angles, with generally distinct left- and right-handed hierarchies when $p_{12}\neq p_{21}$.
The observable mixing from this block is governed by the column suppression (left rotation),
and in general
\begin{equation}
\theta_{ij}^L \sim B^{-(p_{ij}-p_{jj})/9},
\end{equation}
ensuring that off-diagonal elements are suppressed relative to diagonal entries.
Large leptonic mixing emerges when the neutrino sector provides comparable
rotations that interfere with the small charged-lepton rotations.

\subsection*{Tribimaximal limit}

The FX parameterization makes the role of its $12$ rotation
$\theta_\ell$ especially transparent.
Setting $\theta_\ell=0$ gives
$U_{e3}=-s_\ell\,e^{i\phi_{\rm FX}}\sin\theta=0$, and the PMNS matrix
reduces to
\begin{equation}
U_{\rm PMNS}^{(\theta_\ell=0)}=
\begin{pmatrix}
c_\nu & s_\nu & 0\\[4pt]
-e^{i\phi}\,c\,s_\nu & e^{i\phi}\,c\,c_\nu & e^{i\phi}\,s\\[4pt]
s\,s_\nu & -s\,c_\nu & c
\end{pmatrix},
\label{eq:TBM-FX}
\end{equation}
where $c\equiv\cos\theta$, $s\equiv\sin\theta$, and the shorthand
$c_\nu,s_\nu$ stands for $\cos\theta_\nu,\sin\theta_\nu$.
Substituting the tribimaximal (TBM) values~\cite{HarrisonPerkinsScott}
$\theta_\nu=\arctan(1/\!\sqrt{2})\approx 35.26^\circ$ and $\theta=\pi/4$
reproduces the familiar TBM pattern
\begin{equation}
|U_{\rm TBM}|=
\begin{pmatrix}
\sqrt{2/3} & 1/\sqrt{3} & 0\\[3pt]
1/\sqrt{6} & 1/\sqrt{3} & 1/\sqrt{2}\\[3pt]
1/\sqrt{6} & 1/\sqrt{3} & 1/\sqrt{2}
\end{pmatrix}.
\label{eq:TBM-magnitudes}
\end{equation}
In this limit the phase $\phi_{\rm FX}$ is unobservable: it multiplies the
entire second row and can be absorbed by a muon-field redefinition,
giving $J_{\rm CP}=s_\ell c_\ell\cdots\sin\phi_{\rm FX}=0$.

Three consequences are immediate:
(i) exact TBM is the $\theta_\ell=0$ point of the FX decomposition,
parameterized by $(\theta_\nu,\theta)$ alone;
(ii) in this decomposition all departures from TBM (nonzero $\theta_{13}$,
atmospheric octant displacement, and the Dirac CP phase) appear together
when $\theta_\ell\neq 0$;
(iii) $\theta_\ell$ is an effective angle, not the physical charged-lepton
rotation: the identity $\sin\theta_\ell\sin\theta=\sin\theta_{13}$ fixes
$\theta_\ell\simeq\sqrt2\,\theta_{13}\approx12^\circ$ from the
neutrino-sector reactor amplitude $s^\nu_{13}\sim\e$ of
Sec.~\ref{sec:mutau}, while the physical charged-lepton rotations,
$\theta^e_{ij}\sim\e^{17/9},\e^{2},\e^{20/9}$ of
Sec.~\ref{sec:charged-lepton}, enter the FX angles and phase only through
the interference in $U_{e3}$ that shifts $\delta$ from $\delta_\nu$ and
selects the octant. The same small parameter $\e$ thus sets the reactor
angle in the neutrino sector and, through the charged-lepton hierarchy, the
size of the octant and Dirac-phase displacements, producing the two-branch
correlation analyzed in Sec.~\ref{sec:scan_corr}.

\section{Single-\texorpdfstring{$B$}{B} Framework}

We work with
\begin{equation}
\e \equiv \frac{1}{\Bpar},\qquad
\Bpar=\frac{75}{14}\simeq 5.357,\qquad
\e\simeq 0.1867.
\end{equation}
Lepton mixing factorizes as in Eq.~\eqref{eq:PMNS-factorize}.

\section{Charged-Lepton Sector: Analytic Eigenvectors}
\label{sec:charged-lepton}

We adopt the symmetric texture
\begin{equation}
p^e=\frac{1}{18}
\begin{pmatrix}
87&64&40\\
64&30&36\\
40&36&0
\end{pmatrix},
\qquad
(\Ye)_{ij}=c^e_{ij}\e^{p^e_{ij}}.
\label{eq:pe-texture}
\end{equation}

\noindent
\emph{Charges, masses, and mixing.}
Each Yukawa entry arises from a messenger chain carrying $p^e_{ij}$ powers of
$\langle\Phi\rangle/\Lambda$; the diagonal exponents set the charged-lepton mass
hierarchy, while the observable mixing comes from the left-handed column
differences $p^e_{ij}-p^e_{jj}$, giving $\theta^e_{12}\sim\e^{17/9}$,
$\theta^e_{23}\sim\e^{2}$, and $\theta^e_{13}\sim\e^{20/9}$.
These exponents are not those of a single composed chain.
Composing the $12$ and $23$ steps would give
$\theta^e_{13}\sim\e^{17/9+18/9}=\e^{35/9}$, whereas the texture has the milder
$\e^{20/9}$, so the $(1,3)$ entry is set by a separate, shorter chain.
We therefore take Eq.~\eqref{eq:pe-texture} as a phenomenological texture fixed
by the observed masses and mixing; the off-diagonal suppression it requires
beyond additivity cannot come from the messenger content alone and corresponds
to a generation-dependent selection rule, a ``wrinkle'' in the language of
Ref.~\cite{AsadiWrinkles}, for which an explicit example is given in
Appendix~\ref{app:charges-scan}.

The large atmospheric and solar mixing angles are not produced by this
charged-lepton structure; they arise in the neutrino sector from the flat
$\mathcal{O}(1)$ texture of Eq.~\eqref{eq:pnu-texture} together with
mu--tau reflection symmetry, developed in Sec.~\ref{sec:mutau}.
The relation of this flat texture to the hierarchical Weinberg-operator
assignment is given in Appendix~\ref{app:companion}.

\paragraph*{Non-uniqueness and minimal symmetric completion.}
The left-handed power counting is controlled by column differences
$p^{eL}_{ij}\equiv p^e_{ij}-p^e_{jj}$ (since $(U_{eL})_{ij}\sim \epsilon^{\,p^e_{ij}-p^e_{jj}}$).
Therefore, for a fixed $p^{eL}$ the most general (possibly asymmetric) texture is
\begin{equation}
p^e_{ij}=p^{eL}_{ij}+d_j,\qquad d_j\equiv p^e_{jj},
\end{equation}
i.e.\ the three diagonal exponents $d_j$ parameterize the remaining freedom and determine the
right-handed differences $p^{eR}_{ij}=p^e_{ij}-p^e_{ii}=p^{eL}_{ij}+d_j-d_i$.
Imposing the additional condition $p^e=p^{e\,T}$ fixes the diagonal differences uniquely
(up to an overall shift) via $d_j-d_i=p^{eL}_{ji}-p^{eL}_{ij}$, yielding the symmetric
matrix used in this paper as the minimal symmetric completion of the chosen $U_{eL}$ power counting.

The diagonal entries of Eq.~\eqref{eq:pe-texture} ($p^e_{11}=\tfrac{29}{6}$,
$p^e_{22}=\tfrac{5}{3}$, $p^e_{33}=0$) fix the charged-lepton mass hierarchy:
\begin{equation}
m_e : m_\mu : m_\tau \;\sim\;
c_e\,\e^{29/6} \;:\; c_\mu\,\e^{5/3} \;:\; 1,
\label{eq:me-hierarchy}
\end{equation}
where $c_e,\,c_\mu=\mathcal{O}(1)$.
Using $\overline{\rm MS}$ masses at $M_Z$~\cite{Barger2025bfnb},
$\e^{29/6}\simeq 3.0\times 10^{-4}$ and
$\e^{5/3}\simeq 6.1\times 10^{-2}$, giving
$c_e\simeq 0.93$ and $c_\mu\simeq 0.97$.
The muon coefficient $c_\mu$ is fixed by the $\mathcal{O}(1)$ entries
in the $(2,2)$ and $(2,3)$ sub-block of the Yukawa matrix.
The off-diagonal entry $(\Ye)_{23}\sim\e^2$ is $\mathcal{O}(\e^{1/3})\approx 57\%$
of the diagonal $(\Ye)_{22}\sim\e^{5/3}$ in amplitude, yet it enters the
muon mass eigenvalue only at second order, shifting it by
$\mathcal{O}(\e^{7/3})\approx 2\%$; this is why $c_\mu$ remains near unity.

To leading nontrivial order, the orthonormal eigenvectors of $\Ye\Ye^\dagger$
are
\begin{align}
v_\tau &\simeq (0,\,0,\,1)^T
+ \mathcal{O}(\e^2),
\\
v_\mu &\simeq (0,\,1,\,-\e^2)^T
+ \mathcal{O}(\e^{17/9}),
\\
v_e &\simeq (1,\,-\e^{17/9},\,-\e^{20/9})^T
+ \mathcal{O}(\e^{2}).
\end{align}
Thus
\begin{equation}
\Ue \simeq
\begin{pmatrix}
1 & \e^{17/9} & \e^{20/9}\\
-\e^{17/9} & 1 & \e^2\\
-\e^{20/9} & -\e^2 & 1
\end{pmatrix}.
\end{equation}

\section{Neutrino Sector: Mass Texture and Near-TBM Mixing}

\subsection{\texorpdfstring{$B$}{B}-lattice mass hierarchy}

For normal ordering the $B$-lattice assigns
\begin{equation}
m_3:m_2:m_1\sim 1:\e:\e^2,
\end{equation}
giving $m_3\simeq 50~\mathrm{meV}$, $m_2\simeq 9~\mathrm{meV}$,
$m_1\simeq 2~\mathrm{meV}$.
This hierarchy is a direct consequence of the $B$-lattice power counting
and holds independently of the $\mathcal{O}(1)$ structure of the neutrino
mass matrix.
The ratio $m_2:m_3$ fixes the solar splitting only up to this
$\mathcal{O}(1)$ freedom: the pure scaling $m_2=\e\,m_3\simeq 9.3~\mathrm{meV}$
gives $\Delta m^2_{21}$ about $13\%$ high, so the measured value is matched by
a coefficient $\approx 0.9$ on $m_2$ (equivalently $m_2\simeq 8.6~\mathrm{meV}$,
the value used in Sec.~\ref{sec:0nubb}), rather than by $\e$ exactly.

\subsection{Mu--tau reflection symmetry and the \texorpdfstring{$\mathcal{O}(1)$}{O(1)} structure}
\label{sec:mutau}

The $B$-lattice hierarchy alone does not determine the neutrino mixing
pattern.  Large atmospheric and solar angles require additional structure
in the $\mathcal{O}(1)$ coefficients of the effective mass matrix
\begin{equation}
(\mnu)_{ij} = c^\nu_{ij}\,\e^{\,p^\nu_{ij}}\,m_0,
\label{eq:mnu-texture}
\end{equation}
where $m_0$ sets the overall neutrino mass scale.
The FN charge assignment $p^\nu_{L_1}\gg p^\nu_{L_2}=p^\nu_{L_3}$
produces the exponent structure
\begin{equation}
p^\nu=\frac{1}{9}
\begin{pmatrix}
9 & 9 & 9\\
9 & 0 & 0\\
9 & 0 & 0
\end{pmatrix},
\label{eq:pnu-texture}
\end{equation}
which suppresses the first-generation couplings by $\mathcal{O}(\e)$
relative to the $23$ block.  However, the
$\mathcal{O}(1)$ entries must satisfy \emph{mu--tau reflection
symmetry}~\cite{HarrisonScott,GrimusLavoura,MuTauReview}, the exchange
$\nu_\mu\leftrightarrow\nu_\tau$ combined with CP, under which, in the
lattice basis,
\begin{equation}
c^\nu_{22}=c^{\nu*}_{33},
\qquad
c^\nu_{12}=c^{\nu*}_{13},
\qquad
c^\nu_{11},\;c^\nu_{23}\in\mathbb{R},
\label{eq:mutau-coeff}
\end{equation}
together with a near-maximal ratio
$\mathrm{Re}\,c^\nu_{23}/\mathrm{Re}\,c^\nu_{22}\simeq -1$.

The matrix that diagonalizes a mass matrix obeying
Eq.~\eqref{eq:mutau-coeff} has $\theta^\nu_{23}=45^\circ$ and
$\delta_\nu=\pm90^\circ$ exactly, Majorana phases $0$ or $\pi$, and an
unconstrained reactor angle~\cite{HarrisonScott,GrimusLavoura}.
The real limit, $c^\nu_{22}=c^\nu_{33}$ and $c^\nu_{12}=c^\nu_{13}$ real,
is the mu--tau exchange symmetry $Z_2^{\mu\tau}$ under which
$L_\mu\leftrightarrow L_\tau$~\cite{Lam}; it forces $\theta^\nu_{13}=0$
and admits no CP phase.
In the language of residual symmetries, reflection is the
$Z_2\times{\rm CP}$ symmetry of the neutrino sector; it is realized, for
example, by the flavor group $S_4$ combined with a generalized CP
transformation~\cite{FHZ,HLS}, in which case the solar angle is also fixed
toward its tribimaximal value~\cite{A4Review,MaRajasekaran,AltarelliFeruglio}.
The symmetry is a property of the $\mathcal{O}(1)$ coefficients
$c^\nu_{ij}$ of the Weinberg operator in the lattice basis.
The flavon enters the flat texture of Eq.~\eqref{eq:pnu-texture} only
through the factor $\e\,e^{i\phi_0}$ on the first row and column, with
integer exponent, which is real for $\phi_0=\pi$, the value adopted in
Sec.~\ref{sec:phase-alignment}; at $\phi_0=\pi$ the mass matrix is
therefore reflection symmetric exactly when its coefficients are, with
$\delta_\nu=270^\circ$, and a departure of $\phi_0$ from $\pi$ shifts
$\delta_\nu$ by approximately $\phi_0-\pi$.
The real parts of the coefficient splittings,
$\mathrm{Re}(c^\nu_{12}-c^\nu_{13})$ and $c^\nu_{22}-c^\nu_{33}$, shift
$\delta_\nu$ by a few degrees.
The scan of Sec.~\ref{sec:scan_corr} allows $\delta_\nu=270^\circ\pm16^\circ$
for these breaking effects, corresponding to $|\phi_0-\pi|\lesssim15^\circ$.
The realization of the generalized CP symmetry in the ultraviolet, and its
breaking by the charged-lepton textures, are not addressed here.

With this input, the leading-order
effective mass matrix takes the form
\begin{equation}
\frac{\mnu}{m_3}\;\simeq\;
\begin{pmatrix}
a\,\e & r\,\e & r^{*}\e\\[2pt]
r\,\e & \;\;\;\tfrac12 & -\tfrac12 \\[2pt]
r^{*}\e & -\tfrac12 & \;\;\;\tfrac12
\end{pmatrix}\!,\qquad a\in\mathbb{R},\ r\in\mathbb{C},
\label{eq:mnu-LO}
\end{equation}
the form of Refs.~\cite{HarrisonScott,GrimusLavoura},
whose $23$ block has eigenvalues $0$ and $1$ (in units of $m_3$), giving
maximal atmospheric mixing exactly, while the first row mixes via the
$m_2$ eigenvalue to produce a near-tribimaximal solar
angle~\cite{HarrisonPerkinsScott} and, through $\mathrm{Im}\,r$, the
reactor angle and the maximal neutrino phase of the next subsection.

\subsection{The reactor angle and the neutrino phase from the reflection-symmetric first row}

Exact mu--tau \emph{exchange} forces $\theta_{13}=0$, which is excluded by
experiment; reflection does not.  In Eq.~\eqref{eq:mnu-LO} the first-row
difference
\begin{equation}
(\mnu)_{12}-(\mnu)_{13}=2i\,\mathrm{Im}(r)\,\e\,m_3
\label{eq:mutau-breaking}
\end{equation}
is purely imaginary and of order $\e\,m_3$; it generates
$\theta_{13}^\nu\sim\e$, numerically $\approx 0.15$~rad for
$|\mathrm{Im}\,r|\approx 0.7$, without breaking the symmetry, and fixes the
neutrino-sector phase at $\delta_\nu=270^\circ$ (or $90^\circ$, the
CP-conjugate choice of $\mathrm{sgn}\,\mathrm{Im}\,r$).
The reflection symmetry is broken in two places: in the PMNS matrix by the
charged-lepton rotations $\Ue$, which are the source of the octant and
Dirac-phase displacements analyzed in Sec.~\ref{sec:scan_corr}, and in the
neutrino sector by the real parts of the coefficient splittings and by a
departure of the flavon phase $\phi_0$ from $\pi$
(Sec.~\ref{sec:mutau}), which shift $\delta_\nu$ by the amounts quoted
there.  A high-scale completion of the Weinberg operator, for example the
type-I seesaw~\cite{Minkowski,Yanagida,GellMannRamond,MohapatraSenjanovic},
is not required for any of these statements.

\subsection{Resulting \texorpdfstring{$\Unu$}{Unu} and summary}

The combined hierarchy plus reflection structure yields a
near-tribimaximal diagonalization matrix
\begin{equation}
\Unu\simeq
\begin{pmatrix}
\frac{2}{\sqrt6}&\frac{1}{\sqrt3}&\mathcal{O}(\e)\\[4pt]
-\frac{1}{\sqrt6}&\frac{1}{\sqrt3}&\frac{1}{\sqrt2}\\[4pt]
-\frac{1}{\sqrt6}&\frac{1}{\sqrt3}&-\frac{1}{\sqrt2}
\end{pmatrix},
\end{equation}
whose columns are the eigenvectors
\begin{align}
u_3 &\simeq \frac{1}{\sqrt2}(0,\,1,\,-1)^T,
\\
u_2 &\simeq \frac{1}{\sqrt3}(1,\,1,\,1)^T,
\\
u_1 &\simeq \frac{1}{\sqrt6}(2,\,-1,\,-1)^T,
\end{align}
up to the $\mathcal{O}(\e)$ first-row entries that generate $\theta_{13}^\nu\sim\e$
with $\delta_\nu=270^\circ$.

Thus the neutrino sector draws on two distinct ingredients: the $B$-lattice
hierarchy for the mass eigenvalues and the first-generation suppression,
and mu--tau reflection symmetry, a $Z_2\times{\rm CP}$ residual symmetry
realizable within $S_4$ with generalized CP, for the $\mathcal{O}(1)$
structure of the $23$ block and the first row.
The charged-lepton sector, by contrast, requires only FN power counting.
The interplay between these two sectors (neutrino-dominant large angles
corrected by small charged-lepton rotations) is the mechanism that
produces the predictive octant--$\delta$ correlation analyzed in
Sec.~\ref{sec:scan_corr}.

\section{Numerical Diagonalization and PMNS Matrix}

We adopt a representative benchmark in the NO-I region identified by the scan
(Sec.~\ref{sec:scan_corr}): neutrino-sector angles
$\theta_{12}^\nu=34^\circ$, $\theta_{23}^\nu=45^\circ$,
$\theta_{13}^\nu=0.158$~rad, phase $\delta_\nu=270^\circ$ (mu--tau
reflection), and $\mathcal{O}(1)$ charged-lepton coefficients
$c_{12}^e=0.8$, $c_{23}^e=1.08$, $c_{13}^e=0.8$, with aligned phases
$\phi_{12}^e=\phi_{23}^e=\phi_{13}^e=20^\circ$.
Numerical diagonalization gives the PMNS magnitudes in Table~\ref{tab:PMNSnum}
and the PDG observables $\sin^2\theta_{12}=0.303$, $\sin^2\theta_{23}=0.465$,
$\sin^2\theta_{13}=0.0223$, and $\delta\simeq255^\circ$.

\begin{table*}[t]
\centering
\setlength{\tabcolsep}{4pt}
\renewcommand{\arraystretch}{1.05}
\caption{Numerical PMNS matrix (absolute values) for the representative
NO-I benchmark described in the text ($\delta_\nu=270^\circ$ from mu--tau reflection).}
\label{tab:PMNSnum}
\begin{tabular}{ccc}
\toprule
$|U_{e1}|$ & $|U_{e2}|$ & $|U_{e3}|$\\
\midrule
0.83 & 0.54 & 0.15\\
\midrule
$|U_{\mu1}|$ & $|U_{\mu2}|$ & $|U_{\mu3}|$\\
\midrule
0.39 & 0.63 & 0.67\\
\midrule
$|U_{\tau1}|$ & $|U_{\tau2}|$ & $|U_{\tau3}|$\\
\midrule
0.41 & 0.56 & 0.72\\
\bottomrule
\end{tabular}
\end{table*}

The leptonic Jarlskog invariant is
\begin{equation}
\JCP=\mathrm{Im}[U_{e1}U_{\mu2}U_{e2}^*U_{\mu1}^*]
\simeq -0.032,
\end{equation}
consistent with $\JCP\sim s_{12}c_{12}\,s_{23}c_{23}\,s_{13}c_{13}^2\sin\delta$.
This benchmark value $\simeq-0.032$ sits close to the $2\sigma$ lower-octant
scan mean $\JCP\simeq-0.033$ (Table~\ref{tab:branch-averages}); the scan
dispersion of $\delta$ is tight ($\sigma_\delta\simeq7^\circ$,
Sec.~\ref{sec:scan_corr}), so the branch-averaged $J_{\rm CP}$ stays near the
benchmark rather than being washed out.

\section{NuFIT 6.0 Comparison}

Using NuFIT~6.0~\cite{NuFIT60} (NO) central values, we compute illustrative pulls
\begin{equation}
\chi^2=\sum_i\frac{(x_i-x_i^{\rm bf})^2}{\sigma_i^2}.
\end{equation}

\begin{table*}[t]
\centering
\caption{Pulls (normal ordering) for the NO-I benchmark of
Table~\ref{tab:PMNSnum}. Model entries are the benchmark PDG observables.
The three angles are referenced to the NuFIT~6.0 NO best fit (with SK
atmospheric data, lower-octant $\theta_{23}$): $\sin^2\theta_{12}=0.307$,
$\sin^2\theta_{23}=0.470$, $\sin^2\theta_{13}=0.0222$. The phase $\delta$ is
compared to the T2K normal-ordering preference for near-maximal CP
violation~\cite{T2K}; the NuFIT global $\delta$ is pulled toward CP
conservation by the reactor and NOvA tension, whereas the benchmark
$\delta=255^\circ$ lies within the T2K $1\sigma$ region.}
\label{tab:pulls}
\begin{tabular}{ccc}
\toprule
Observable & Model & Pull\\
\midrule
$\sin^2\theta_{12}$ & 0.303 & $-0.4\sigma$\\
$\sin^2\theta_{23}$ & 0.465 & $-0.4\sigma$\\
$\sin^2\theta_{13}$ & 0.0223 & $+0.1\sigma$\\
$\delta$ & $255^\circ$ & $-0.7\sigma$\\
\bottomrule
\end{tabular}
\end{table*}

The per-observable pulls are collected in Table~\ref{tab:pulls}; the total
$\chi^2\simeq 0.7$ (with $\delta$ referenced to the T2K near-maximal
preference), confirming that the NO-I region of the coefficient scan is
compatible with current data.

\section{Discrete Symmetry and Anomaly Consistency}
\label{sec:discrete}

The lattice textures can be enforced by a discrete $Z_N$ symmetry with
$N=18$ or $54$, consistent with the lattice interpretation.
Assigning FN charges
\begin{equation}
Q(L_i)+Q(E^c_j)\equiv p^e_{ij}\pmod N,
\end{equation}
the Yukawa operator
\begin{equation}
L_i H_d E^c_j\left(\frac{\phi}{M}\right)^{p^e_{ij}}
\end{equation}
is invariant.
Discrete gauge anomaly cancellation follows the Ib\'a\~nez--Ross conditions~\cite{IbanezRoss},
which can be satisfied generation by generation for $Z_{18}$ or $Z_{54}$
with the above charge assignments.
\section{Majorana Phases and Neutrinoless Double Beta Decay}
\label{sec:0nubb}

If neutrinos are Majorana, the PMNS matrix contains two additional physical
phases beyond the Dirac phase.  In the PDG convention one writes
\begin{equation}
\UPMNS^{\rm (PDG)} \;=\;
\widetilde U(\theta_{12},\theta_{23},\theta_{13},\delta)\,
\diag\!\left(1,\;e^{i\alpha_{21}/2},\;e^{i\alpha_{31}/2}\right),
\label{eq:PMNS-PDG-Maj}
\end{equation}
where $\widetilde U$ carries the three mixing angles and Dirac phase $\delta$,
and $(\alpha_{21},\alpha_{31})$ are the Majorana phases.

The neutrinoless double beta decay amplitude is controlled by the effective
Majorana mass
\begin{equation}
\begin{split}
m_{\beta\beta}&\;\equiv\;
\left|\sum_{i=1}^3 (\UPMNS)_{ei}^2\,m_i\right|
\\
&=\bigl|\,
 m_1\,c_{12}^2 c_{13}^2
+ m_2\,s_{12}^2 c_{13}^2\,e^{i\alpha_{21}}
\\
&\qquad
+ m_3\,s_{13}^2\,e^{i(\alpha_{31}-2\delta)}
\bigr|,
\end{split}
\label{eq:mbb-def}
\end{equation}
with $s_{ij}\equiv\sin\theta_{ij}$, $c_{ij}\equiv\cos\theta_{ij}$.

\subsection{Single-\texorpdfstring{$\Bpar$}{B} scaling for \texorpdfstring{$m_{\beta\beta}$}{mbetabeta} (normal ordering)}
In our benchmark normal ordering scaling
\begin{equation}
m_3:m_2:m_1 \sim 1:\e:\e^2,
\label{eq:mi-scaling}
\end{equation}
and with $s_{13}^2\sim\e^2$ and $c_{13}\simeq 1$,
the three contributions to Eq.~\eqref{eq:mbb-def} scale as
\begin{equation}
m_{\beta\beta} \sim \mathcal{O}(m_3)\times
\max\!\left(c_{12}^2\e^2,\;s_{12}^2\e,\;\e^2\right)
\sim s_{12}^2\,\e\,m_3,
\label{eq:mbb-scaling}
\end{equation}
where the dominant term arises from the $m_2$ contribution modulated by $s_{12}^2\simeq 0.30$.
Thus, in the absence of tuned cancellations, the single-$\Bpar$ framework predicts
\begin{equation}
m_{\beta\beta} \;\sim\; \frac{s_{12}^2\,m_3}{\Bpar}.
\label{eq:mbb-Bpower}
\end{equation}

\subsection{Illustrative numerical range}
Adopting a representative normal-ordering spectrum with $m_3\simeq 50~\mathrm{meV}$
and $m_2\simeq 8.6~\mathrm{meV}$, the three terms in Eq.~\eqref{eq:mbb-def} evaluate to
approximately $1.2$, $2.5$, and $1.1~\mathrm{meV}$ respectively, giving
\begin{equation}
m_{\beta\beta}\sim 1\text{--}5~\mathrm{meV},
\label{eq:mbb-numerical}
\end{equation}
where the spread is the envelope over the Majorana phases in
Eq.~\eqref{eq:mbb-def}.  The reflection symmetry of
Sec.~\ref{sec:mutau} fixes the Majorana phases to $0$ or $\pi$ at leading
order, so that within this envelope $m_{\beta\beta}$ takes one of a
discrete set of values, $(\alpha_{21},\alpha_{31})=(0,0),(\pi,0),(0,\pi),(\pi,\pi)$
giving $m_{\beta\beta}\simeq 2.9$, $2.3$, $4.8$, and $0.7~\mathrm{meV}$ at the
benchmark of Sec.~\ref{sec:scan_corr}, up to $\mathcal{O}(\e)$ corrections.
This is below current experimental limits but within reach of next-generation
experiments aiming at the few-meV regime~\cite{0nubbReview}.
\section{Phenomenology of the Two Normal-Ordering (NO) Branches}
\label{sec:twoNO}

Global fits to oscillation data (NuFIT 6.0~\cite{NuFIT60}) have two
normal-ordering (NO) local minima, at $\sin^2\theta_{23}\approx0.47$ and
$\approx0.56$, differing in the atmospheric octant.  The single-$\Bpar$
framework has two branches, from the two signs of the charged-lepton
interference in $\UPMNS = \Ue^\dagger \Unu$ [Eq.~\eqref{eq:PMNS-factorize}],
which lie slightly below and slightly above maximal mixing,
$\theta_{23}\approx43$--$44^\circ$ and $46$--$47^\circ$
($\sin^2\theta_{23}\approx0.47$ and $0.52$).  The lower branch coincides with
the lower-octant minimum; the upper branch shares the octant of the upper
minimum but not its value.  Reaching $\theta_{23}=48.4^\circ$
($\sin^2\theta_{23}=0.56$) requires $\theta^e_{23}\cos\phi^e_{23}=-3.4^\circ$,
i.e.\ $c^e_{23}\simeq1.7$ at full anti-alignment, outside the range
$[0.5,1.5]$ of $\mathcal{O}(1)$ coefficients adopted in the scan, whose
largest value is $48.3^\circ$.  A confirmed $\sin^2\theta_{23}\approx0.56$
would therefore strain the framework rather than realize its upper branch,
and a value above $0.56$ would exclude it (Sec.~\ref{sec:conclusions}).

\subsection{NO-I: Lower-Octant Branch}

In the first branch, $\theta_{23}<45^\circ$ and the Dirac phase is
in the range $\delta\sim 240^\circ$--$280^\circ$ (centered near $260^\circ$).
In our framework this arises when the cancellation in
\begin{equation}
U_{e3} \simeq \theta_{13}^\nu - \theta_{12}^e \sin\theta_{23}^\nu e^{i\phi}
\end{equation}
is moderate, so that $|U_{e3}|\sim \e$ without strong fine tuning.

Predictions:
\begin{itemize}
\item $\theta_{23}$ slightly below maximal.  The maximal neutrino value
$\theta^\nu_{23}=45^\circ$ is shifted by the charged-lepton $23$ rotation,
$\theta_{23}\simeq\theta^\nu_{23}-\theta^e_{23}\cos\phi^e_{23}$
[Eq.~\eqref{eq:theta23-theorem}]; on this branch $\cos\phi^e_{23}>0$, so the
$\mathcal{O}(\e^2)$ rotation $\theta^e_{23}\approx2^\circ$ subtracts and leaves
$\theta_{23}$ about $2^\circ$ below maximal.
\item $\delta$ in the third quadrant, displaced $\sim\!10$--$15^\circ$ below
the maximal neutrino phase $270^\circ$ by the charged-lepton interference.
\item $m_{\beta\beta}\simeq 3~\mathrm{meV}$ (for vanishing Majorana phases) and $J_{\rm CP}\simeq -0.033$,
essentially identical to NO-II (see Sec.~\ref{sec:scan_corr}).
\end{itemize}

\subsection{NO-II: Upper-Octant Branch}

In the second branch, $\theta_{23}$ lies $1$--$2^\circ$ above maximal and
$\delta$ shifts upward to $\sim\!275^\circ$, above $270^\circ$.  This
corresponds to the opposite sign of $\cos\phi^e_{23}$ in
Eq.~\eqref{eq:theta23-theorem}; in the strict single-source limit it
requires the charged-lepton phases to be displaced from their $\phi_0=\pi$
values (Sec.~\ref{sec:phase-alignment}).

Predictions:
\begin{itemize}
\item $\theta_{23}$ in the upper octant.  Here the relative phase has
$\cos\phi^e_{23}<0$, so the same charged-lepton $23$ rotation
$\theta^e_{23}\approx2^\circ$ adds in Eq.~\eqref{eq:theta23-theorem} and leaves
$\theta_{23}$ about $2^\circ$ above maximal.
\item $\delta$ displaced upward by $\mathcal{O}(15^\circ)$ relative to NO-I,
to the opposite side of $270^\circ$.
\item $m_{\beta\beta}$ and $|J_{\rm CP}|$ essentially unchanged from NO-I;
the octant flip leaves these observables invariant (Table~\ref{tab:NO-summary}).
\end{itemize}

\subsection{Distinguishing Observables}

The two branches can be distinguished experimentally through:
\begin{enumerate}
\item Precision measurement of $\theta_{23}$ octant at DUNE~\cite{DUNE}/Hyper-K~\cite{HyperK}/IceCube~\cite{IceCube}, building on current constraints from Super-K~\cite{SuperK}, NOvA~\cite{NOvA}, and T2K~\cite{T2K}.
T2K has provided some of the strongest existing constraints on $\delta$ and $\theta_{23}$,
including hints for $\delta\sim 270^\circ$ and some preference for the upper octant.
\item Improved determination of $\delta$.
\item Combined fits of $\delta$ and $\theta_{23}$ correlations from DUNE, Hyper-K, IceCube~\cite{IceCube}, JUNO~\cite{JUNO}, and T2K.
\end{enumerate}
Because $J_{\rm CP}$ and $m_{\beta\beta}$ are nearly identical between the two branches,
neither $0\nu\beta\beta$ searches~\cite{0nubbReview} nor
CP-violation measurements alone can discriminate between NO-I and NO-II;
the octant of $\theta_{23}$ and its correlated shift in $\delta$ remain
the definitive discriminators.
Table~\ref{tab:NO-summary} collects the key properties of the two branches.

\begin{table*}[t]
\centering
\caption{Summary comparison of the two normal-ordering branches.
Here LO and HO denote the lower and upper octant,
$\theta_{23}<45^\circ$ and $\theta_{23}>45^\circ$.
NO-I values of $\theta_{23}$ and $\delta$ are those of the benchmark of
Table~\ref{tab:PMNSnum}; NO-II values are the $2\sigma$ scan medians of
Table~\ref{tab:branch-averages} (both used as FX benchmarks; see Table~\ref{tab:FX-fit});
$J_{\rm CP}$ and $m_{\beta\beta}$ are the $2\sigma$ branch means
from Table~\ref{tab:branch-averages} at $\delta_\nu=270^\circ$.
The last column indicates whether the observable
discriminates between the two branches.}
\label{tab:NO-summary}
\setlength{\tabcolsep}{4pt}
\renewcommand{\arraystretch}{1.1}
\begin{tabular}{lccc}
\toprule
Observable & NO-I (LO) & NO-II (HO) & Discrim.?\\
\midrule
$\theta_{23}$ & $43.0^\circ$ & $46.2^\circ$ & \checkmark \\
$\delta$ & $255^\circ$ & $276^\circ$ & \checkmark \\
\midrule
$\theta_\ell$ (FX) & $12.5^\circ$ & $11.8^\circ$ & weak \\
$\theta$ (FX) & $43.7^\circ$ & $46.8^\circ$ & \checkmark \\
$\theta_\nu$ (FX) & $36.3^\circ$ & $33.1^\circ$ & weak \\
$\phi_{\rm FX}$ & $-69^\circ$ & $-88^\circ$ & \checkmark \\
\midrule
$J_{\rm CP}$ & $-0.033$ & $-0.033$ & $\times$ \\
$m_{\beta\beta}$ [meV] & $2.8$ & $2.9$ & $\times$ \\
\midrule
Theory prior & favored & disfavored & n/a \\
\bottomrule
\end{tabular}
\end{table*}

Within the lattice-flavon framework, the two NO branches therefore map onto
distinct interference phases between $\Ue$ and $\Unu$, making the correlation
between $\theta_{23}$ and $\delta$ a sharp test of the single-$\Bpar$
organizing principle.

\begin{table*}[t]
\centering
\caption{Fritzsch--Xing parameters fitted to the PDG mixing angles
for the two normal-ordering branches.
Common inputs: $\sin^2\theta_{12}=0.303$, $\sin^2\theta_{13}=0.0223$.
The atmospheric angle and Dirac phase are the NO-I benchmark values and the
NO-II $2\sigma$ scan medians of Table~\ref{tab:branch-averages}.
The FX parameterization is
$U_{\rm PMNS}^{(\rm FX)} = R_{12}^T(\theta_\ell)\,P_\phi\,R_{23}(\theta)\,R_{12}(\theta_\nu)$.
The PDG $\theta_{23}$ octant ambiguity
($\sin^2\theta_{23}\leftrightarrow 1-\sin^2\theta_{23}$, correlated with a shift in $\delta$)
maps in the FX representation to a correlated shift
$(\theta,\,\phi_{\rm FX})\to(\theta+\Delta\theta,\,\phi_{\rm FX}+\Delta\phi_{\rm FX})$,
with $\Delta\theta\approx +3.2^\circ$ and $\Delta\phi_{\rm FX}\approx -20^\circ$,
while the 12-sector angles $\theta_\ell$ and $\theta_\nu$ shift by less.}
\label{tab:FX-fit}
\begin{tabular}{lcc}
\hline\hline
Parameter & NO-I (LO) & NO-II (HO) \\ \hline
$\sin^2\theta_{23}$ (PDG) & $0.465$ & $0.521$ \\
$\delta$ (PDG) & $255^\circ$ & $276^\circ$ \\ \hline
$\theta_\ell$ & $12.5^\circ$ & $11.8^\circ$ \\
$\theta$ & $43.7^\circ$ & $46.8^\circ$ \\
$\theta_\nu$ & $36.3^\circ$ & $33.1^\circ$ \\
$\phi_{\rm FX}$ & $-69^\circ\;(-0.38\pi)$ & $-88^\circ\;(-0.49\pi)$ \\
\hline\hline
\end{tabular}
\end{table*}

Table~\ref{tab:FX-fit} presents the Fritzsch--Xing rotation
angles and phase obtained by fitting the FX decomposition to the
PDG matrix for each branch.
The FX angle $\theta_\ell\approx 12^\circ$, the effective $12$ rotation
fixed by $\theta_{13}$, is approximately invariant between the two
branches, while $\theta_\nu$ straddles the tribimaximal value
$\theta_\nu^{\rm TBM}\simeq 35.3^\circ$ ($36.3^\circ$ for NO-I,
$33.1^\circ$ for NO-II).
The atmospheric-sector angle $\theta$ shifts
from $43.7^\circ$ in NO-I to $46.8^\circ$ in NO-II,
while the FX phase $\phi_{\rm FX}$ shifts by roughly $-20^\circ$.
Thus the familiar PDG octant ambiguity
$\theta_{23}\leftrightarrow 90^\circ-\theta_{23}$
translates in the FX representation into a correlated
$(\theta,\,\phi_{\rm FX})$ ambiguity: the two branches differ
by $\Delta\theta\approx +3.2^\circ$ and $\Delta\phi_{\rm FX}\approx -20^\circ$,
while the 12-sector angles that encode $\theta_{12}$ and $\theta_{13}$
shift by less.
This structure reflects the fact that, in the FX factorization, the
$R_{23}(\theta)$ rotation and the phase matrix $P_\phi$ jointly
control both $\theta_{23}$ and $\delta$, so a shift in one
necessarily entails a compensating shift in the other.

\subsection{Comparison with the quark-sector FX phase}

In the quark sector, an FX-type fit to the CKM matrix yields
$\phi_{\rm FX}^{\rm (CKM)}\approx 90^\circ$~\cite{Fritzsch,LatticeFlavonQuarkMixing},
i.e.\ maximal CP violation given the angular prefactors.
It is instructive to ask what the corresponding situation is
in the lepton sector.

From Table~\ref{tab:FX-fit}, the leptonic FX phase is
$\phi_{\rm FX}\simeq -69^\circ$ (NO-I) and $-88^\circ$ (NO-II).
Both branches thus realize a near-maximal FX phase, as the quark sector
does, but of opposite sign, near $-90^\circ$ rather than $+90^\circ$, the
sign tracking the negative leptonic Jarlskog ($\sin\delta<0$).
In this parameterization the two sectors are parallel in magnitude and
mirrored in sign.
NO-II lies within $2^\circ$ of $-90^\circ$, NO-I about $21^\circ$ above it;
it is this residual departure, not the value itself, that carries the
model-specific information, through two correlated effects:
\begin{enumerate}
\item \emph{Submaximal leptonic CP violation.}
Since $J_{\rm CP}\propto \sin\phi_{\rm FX}$, the suppression factor
$|\!\sin\phi_{\rm FX}/\sin 90^\circ|$ is $0.93$ for NO-I and
$1.00$ for NO-II.
The leptonic Jarlskog invariant is therefore reduced by
up to $7\%$ relative to its angular bound.

\item \emph{Departure of $\theta_\nu$ from the tribimaximal value.}
Fixing $\phi_{\rm FX}=-90^\circ$ and matching the solar, atmospheric, and
reactor angles yields $\theta_\nu\simeq 33^\circ$, about $2^\circ$ below the
tribimaximal value $\theta_\nu^{\rm TBM}=\arctan(1/\!\sqrt{2})\simeq 35.3^\circ$,
with $\delta\simeq 278^\circ$ (NO-I) and $277^\circ$ (NO-II).
NO-II nearly realizes this point ($\delta=276^\circ$); NO-I, with
$\delta=255^\circ$, departs from it by the $-23^\circ$ that the
charged-lepton interference produces on the lower branch, and the FX fit
absorbs that departure as $\phi_{\rm FX}\simeq-69^\circ$ together with a
$1^\circ$ rise of $\theta_\nu$ above its tribimaximal value.
\end{enumerate}

The quark-sector ``neat $90^\circ$'' thus has a near-maximal counterpart
in the lepton sector, of opposite sign, realized almost exactly on the
upper branch and displaced by $21^\circ$ on the lower.
That displacement is not free.  The neutrino-sector phase is maximal,
$\delta_\nu=270^\circ$ [Eq.~\eqref{eq:delta-theorem}], so the departure of
$\phi_{\rm FX}$ from $-90^\circ$ is the charged-lepton interference shift
of $\delta$ about that maximal value, $-15^\circ$ for NO-I and $+6^\circ$
for NO-II, repackaged by the FX decomposition.
The near-maximality is therefore a parallel to the quark sector and a
consequence of the reflection symmetry, with the branch-dependent
departure a prediction of the charged-lepton sector.
These departures are organized economically by referring the two
branches to a single tribimaximal, maximal-CP reference point;
Sec.~\ref{sec:increments} writes the full lepton mixing as a set of
increments about that point.

\section{Parameter Scan and Analytic Correlations in the Two-NO Picture}
\label{sec:scan_corr}

The two normal-ordering branches (NO-I and NO-II) of the lattice-flavon
framework can be visualized by scanning the $\mathcal{O}(1)$ coefficients and relative
phases entering the small charged-lepton rotations while keeping the neutrino-dominant
angles at their symmetry and best-fit values.
We implement $\UPMNS=\Ue^\dagger\Unu$ with $\Unu$ at $\theta^\nu_{12}=34^\circ$,
$\theta^\nu_{23}=45^\circ$ and $\delta_\nu=270^\circ$ (exact, from mu--tau
reflection), and $\theta^\nu_{13}=0.158$, and $\Ue$ built from the
$B$-lattice power-counting angles $\theta_{12}^e\sim \e^{17/9}$, $\theta_{23}^e\sim \e^{2}$,
$\theta_{13}^e\sim \e^{20/9}$ multiplied by $\mathcal{O}(1)$ coefficients and including relative
phases in the three charged-lepton rotations.
The charged-lepton phases are drawn as $\phi^e_{12}=\mu+g_{12}$ and
$\phi^e_{23}=\mu+g_{23}$, where $g_{12}$ and $g_{23}$ are independent Gaussian
variates of zero mean and width $\sigma=20^\circ$, the single-flavon
misalignment derived below, and the common mean $\mu$ is uniform on
$[0,2\pi)$; no central value is preferred.  The phase $\phi^e_{13}$ is
uniform on $[0,2\pi)$ and the coefficients $c^e_{ij}$ are uniform on
$[0.5,1.5]$.  Because the octant is the sign of $\cos\phi^e_{23}$
[Eq.~\eqref{eq:theta23-theorem}], the uniform mean populates both octants
equally before cuts; the preference for the lower octant found below is
produced by the NuFIT cuts on $\sin^2\theta_{12}$ and $\sin^2\theta_{13}$
and depends on the measure (Appendix~\ref{app:charges-scan}).

\subsection{Phase alignment from the single-flavon structure}
\label{sec:phase-alignment}

The scan assumes approximate alignment of the charged-lepton
$12$ and $23$ rotation phases, $\phi_{12}^e\approx\phi_{23}^e$.
This assumption has two supports: the single-flavon counting below, and
the explicit realization of Appendix~\ref{app:charges-scan}, in which it is
exact.

In a Froggatt--Nielsen model with a single flavon $\Phi$,
each Yukawa entry arises from a messenger chain carrying
$p_{ij}$ powers of $\langle\Phi\rangle/\Lambda$.
If the $\mathcal{O}(1)$ couplings at each messenger vertex are
approximately real, so that CP violation originates dominantly from
the complex phase $\phi_0\equiv\arg\langle\Phi\rangle$ of the
flavon VEV, then
\begin{equation}
Y_{ij} \;\simeq\; |c_{ij}|\,\e^{\,p_{ij}}\,e^{i\,p_{ij}\,\phi_0}.
\end{equation}
For strictly additive exponents $p_{ij}=Q(L_i)+Q(R_j)$ this phase would be
removed by field rephasing, with the flavon phase cancelling between $\Ue$ and
$\Unu$ in $\UPMNS$; the departure from additivity in the charged-lepton sector
(Sec.~\ref{sec:charged-lepton}) is what leaves a physical Dirac phase.
The left-handed rotation that enters $\Ue$ diagonalizes
$Y_e Y_e^\dagger$.  In the hierarchical limit the $ij$ rotation
angle is $\theta_{ij}^e\sim |Y_{ij}/Y_{jj}|$ with phase
\begin{equation}
\phi_{ij}^e \;=\; (p^e_{ij}-p^e_{jj})\,\phi_0.
\end{equation}
For the charged-lepton texture in Eq.~\eqref{eq:pe-texture} this gives
\begin{equation}
\phi_{12}^e = \tfrac{17}{9}\,\phi_0,
\qquad
\phi_{23}^e = \tfrac{18}{9}\,\phi_0 = 2\phi_0,
\end{equation}
so the misalignment is
\begin{equation}
\phi_{12}^e - \phi_{23}^e
\;=\; -\frac{\phi_0}{9}.
\label{eq:phase-misalignment}
\end{equation}
The near-equality $p^e_{12}-p^e_{22}=\tfrac{17}{9}\approx 2=p^e_{23}-p^e_{33}$
ensures that the two rotation phases are automatically
aligned to within $\sim\!\phi_0/9$.
These phases and their misalignment in Eq.~\eqref{eq:phase-misalignment}
are a statement about the model's defining basis (real $\mathcal{O}(1)$
vertex couplings, with CP violation residing in
$\phi_0=\arg\langle\Phi\rangle$); they are not separately invariant under
lepton rephasing, the physical content being carried by the
rephasing-invariant relative phase $\phi_{12}^e-\delta_\nu$ that enters
$U_{e3}$, so the octant--$\delta$ correlation rests on that invariant
rather than on the basis-dependent alignment itself.

For a flavon VEV with $\phi_0\sim\pi$ the
misalignment is $|\phi_{12}^e-\phi_{23}^e|\simeq 20^\circ$; this
motivates the alignment width $\sigma\simeq 20^\circ$ adopted in the
scan.
In the defining basis, with real vertex couplings, the flavon phase
$\phi_0\equiv\arg\langle\Phi\rangle$ is the single source of CP violation
in the charged-lepton Yukawa texture: the rotation phases are
$\phi^e_{ij}=(p^e_{ij}-p^e_{jj})\phi_0$, CP is conserved in that sector
for $\phi_0\to0$, and the fractional lattice exponents make $\phi_0=\pi$
a CP-violating value there, with the largest misalignment,
$\phi_0/9\simeq20^\circ$, consistent with this origin.
This relation presumes that the off-diagonal suppression is itself a power
of the flavon carrying $\phi_0$; in the explicit realization of
Appendix~\ref{app:charges-scan} the suppression is one insertion of a second
spurion $\eta$, the flavon phase is removed by rephasing, and the rotation
phases are $\phi^e_{12}=\phi^e_{23}=\arg\langle\eta\rangle=-\phi^e_{13}$,
with the alignment exact and no link to $\phi_0$.  The default prescription
of the scan, a common uniform mean for $\phi^e_{12}$ and $\phi^e_{23}$, is
this case up to the $20^\circ$ spread; imposing the relation exactly gives
$\theta_{23}=44.0\pm0.6^\circ$, $\delta=257\pm3^\circ$ and
$46.2\pm0.7^\circ$, $272\pm6^\circ$ at exact reflection, with the octant
set by $\mathrm{sgn}\cos\arg\langle\eta\rangle$ and
$N_{\rm LO}/N_{\rm HO}\simeq0.8$.  The strict single-source variant of
Table~\ref{tab:phase-variants} is the opposite limit, in which all phases
trace to $\phi_0$.
Multiple messenger species or complex $\mathcal{O}(1)$ vertex couplings
would scatter the phases around the single-source baseline, but by
amounts of order $\phi_0/9$ per additional phase, preserving
approximate alignment.
The neutrino-sector phase is fixed by the mu--tau reflection of the
$\mathcal{O}(1)$ coefficients, $\delta_\nu=270^\circ$
(Sec.~\ref{sec:mutau}), with the sign a discrete choice; $\phi_0$ enters
the flat neutrino texture only as the factor $e^{i\phi_0}$ on the first
row, real at $\phi_0=\pi$, and a departure from $\pi$ is one of the
breaking effects covered by the band $\delta_\nu=270^\circ\pm16^\circ$.

The $13$ rotation phase satisfies
$\phi_{13}^e = \tfrac{20}{9}\phi_0$, so
$\phi_{13}^e - \phi_{23}^e = \tfrac{2}{9}\phi_0$, a misalignment
twice as large; and the $(1,3)$ entry is set by a separate messenger chain
(Sec.~\ref{sec:charged-lepton}), whose vertex couplings need not share the
phase of the chain that fixes the other two entries.  For these reasons the
scan treats $\phi_{13}^e$ as an independent uniform variable, the
conservative choice; it enters only the subleading
$\theta^e_{13}c^\nu_{23}$ term in $U_{e3}$.
Neither $\delta_\nu$ nor $\phi^e_{13}$ adds an observable Dirac CP-odd
invariant beyond $J_{\rm CP}$, since only rephasing-invariant combinations
such as $\Phi_e=\phi^e_{12}-\delta_\nu$ enter the observables; the
octant--$\delta$ correlation rests on $\Phi_e$, and with
$\delta_\nu=270^\circ$ it reduces to the sign of
$\cos\phi^e_{12}\approx\cos\phi^e_{23}$, the same quantity that selects the
octant.
At $\phi_0=\pi$ the strict single-source charged-lepton phases are
$\phi^e_{23}=2\pi\equiv0^\circ$, $\phi^e_{12}=\tfrac{17}{9}\pi\equiv-20^\circ$,
and $\phi^e_{13}=\tfrac{20}{9}\pi\equiv+20^\circ$.
With the phases set exactly to $(\tfrac{17}{9},2,\tfrac{20}{9})\phi_0$ and
only the magnitudes varied, the solar and reactor cuts accept $\phi_0$
within about $\pm15^\circ$ of $0^\circ$, $90^\circ$, $180^\circ$, and
$270^\circ$, and the octant is $\mathrm{sgn}\cos2\phi_0$: lower at
$\phi_0\simeq0^\circ$ or $180^\circ$ ($\theta_{23}\simeq43^\circ$,
$\delta\simeq255^\circ$), upper at $\phi_0\simeq90^\circ$ or $270^\circ$
($\theta_{23}\simeq47^\circ$, $\delta\simeq281^\circ$), with the sign
relation of Eq.~\eqref{eq:octant-quad} exact
(Table~\ref{tab:phase-variants} in Appendix~\ref{app:charges-scan}).
The former pair is where the flavon factor on the neutrino first row is
real and reflection symmetry holds, the latter where it is $\pm i$ and the
neutrino phase is driven toward CP conservation; in the strict
single-source limit the lower octant and the near-maximal Dirac phase
therefore arise together, and the upper-octant branch requires the phases
to be displaced from their single-source values by the scattering described
above.

\subsection{Scan result: two branches in the \texorpdfstring{$(\theta_{23},\delta)$}{(theta23, delta)} plane}

\sbox{\figscanbox}{%
%
}
\begin{figure}[htbp]
\centering
\resizebox{\columnwidth}{!}{%
\usebox{\figscanbox}%
}
\caption{Scan results in the $(\theta_{23},\delta)$ plane, both octants on a
single axis. The neutrino angles are held at $\theta^\nu_{12}=34^\circ$,
$\theta^\nu_{23}=45^\circ$, $\theta^\nu_{13}=0.158$ with $\delta_\nu=270^\circ$
from mu--tau reflection, while the charged-lepton coefficients and phases are
scanned as described in Appendix~\ref{app:charges-scan}; the plotted points pass
the NuFIT~6.0 $2\sigma$ cuts on $\sin^2\theta_{12}$ and $\sin^2\theta_{13}$.
The lower-octant (LO) and upper-octant (HO) branches are correlated with
$\delta$ below and above $270^\circ$ respectively, so that the points form a
single band of positive slope through $(45^\circ,270^\circ)$ (dashed lines)
rather than two separated populations; the distribution of $\theta_{23}$
itself is shown in Fig.~\ref{fig:theta23-hist}. Shaded bands show the
NuFIT~6.0 $1\sigma$--$3\sigma$ ranges for $\theta_{23}$; stars mark the
branch benchmarks of Table~\ref{tab:NO-summary}.}
\label{fig:scan-delta-theta23}
\end{figure}

\begin{figure}[htbp]
\centering
\resizebox{\columnwidth}{!}{%
\usebox{\fighistbox}%
}
\caption{Distribution of $\theta_{23}$ over both octants for the scan of
Fig.~\ref{fig:scan-delta-theta23}, in $0.25^\circ$ bins. The distribution is
continuous through maximal mixing (dashed line): the two branches are the
two lobes of one distribution, peaked at $\theta_{23}\simeq43.6^\circ$ and
$46.1^\circ$ and separated by a shallow minimum at $45^\circ$ whose density
is two thirds of the lower peak. The lobes are the two signs of
$\cos\phi^e_{23}$ in Eq.~\eqref{eq:theta23-theorem}; before the solar and
reactor cuts they are equally populated, and the cuts retain the lower
octant preferentially ($N_{\rm LO}/N_{\rm HO}=1.35$). The filled bars are
the $\delta_\nu=270^\circ$ scan and the gray outline includes the
reflection-breaking band $\delta_\nu=270^\circ\pm16^\circ$ of
Sec.~\ref{sec:mutau}, which leaves the $\theta_{23}$ distribution unchanged.}
\label{fig:theta23-hist}
\end{figure}

Figure~\ref{fig:scan-delta-theta23} displays the resulting distribution in the plane of the PDG
Dirac phase $\delta$ versus the atmospheric angle $\theta_{23}$, and
Fig.~\ref{fig:theta23-hist} the distribution of $\theta_{23}$ itself.
The lower- and upper-octant branches are the two sides of one continuous
distribution, corresponding to the two signs of the charged-lepton
interference between $\Ue$ and $\Unu$; the distribution of $\theta_{23}$
is continuous through maximal mixing, with lobes at $43.6^\circ$ and
$46.1^\circ$ separated by a shallow minimum at $45^\circ$, and the
$(\theta_{23},\delta)$ points form a single band of positive slope
through $(45^\circ,270^\circ)$.
Before the cuts the two octants are equally populated; the solar and
reactor cuts, acting on the phases through the interference in $U_{e2}$ and
$U_{e3}$, retain the lower octant preferentially,
$N_{\rm LO}/N_{\rm HO}=1.35$.
In this sense the framework carries a mild theoretical prior for NO-I,
which is strengthened by the strict single-source limit of
Sec.~\ref{sec:phase-alignment}, where the lower octant contains the
$\phi_0=\pi$ point.
Current data do not yet resolve the octant:
NuFIT~6.0 finds local minima at both
$\sin^2\theta_{23}\approx 0.47$ and $\approx 0.56$ with
$\Delta\chi^2<4$ between them, and for normal ordering $\delta$
is weakly constrained and consistent with CP conservation
within $1\sigma$~\cite{NuFIT60},
although T2K data favor $\delta\sim 270^\circ$ with some
preference for the upper octant~\cite{T2K}.
The framework's upper branch, $\sin^2\theta_{23}\approx0.52$, sits between
the two minima and below the upper one, so the $1\sigma$ upper-octant
band in Fig.~\ref{fig:scan-delta-theta23} does not overlap it.
Precision measurements of $\theta_{23}$ at
DUNE, Hyper-K, and IceCube~\cite{IceCube} will therefore be decisive in two
ways: a value near $0.56$ would exclude the framework, and a value in the
upper octant near $0.52$ would require $\delta$ above $270^\circ$ along the
correlation band.

The Jarlskog invariant $J_{\rm CP}$ and the effective Majorana mass $m_{\beta\beta}$
are nearly branch-independent, in contrast to $\theta_{23}$ and $\delta$
(Table~\ref{tab:branch-averages}):
both branches yield $J_{\rm CP}\simeq -0.033$ and $m_{\beta\beta}\simeq 3~\mathrm{meV}$
(for vanishing Majorana phases).
This near-degeneracy has two sources.
First, the factor $s_{23}c_{23}=\tfrac12\sin(2\theta_{23})$ in
$J_{\rm CP}$ is stationary at maximal mixing and varies by less than
$0.3\%$ between $\theta_{23}=43.0^\circ$ and $46.2^\circ$.
Second, both branch means lie close to maximal CP violation,
$\delta\simeq 260^\circ$ and $274^\circ$, where $|\sin\delta|\ge0.96$;
the scan dispersion is itself tight
($\sigma_\delta\simeq 7^\circ$--$10^\circ$, set by the charged-lepton interference of
Eq.~\eqref{eq:delta-theorem}), so each branch's $J_{\rm CP}$ is
well-localized rather than smeared.
The two effects leave $J_{\rm CP}\simeq -0.033$ in both octants
($-0.0326$ and $-0.0333$).
Similarly, $m_{\beta\beta}$ is dominated by the $m_2$ term with
coefficient $|U_{e2}|^2\simeq s_{12}^2 c_{13}^2$, which is insensitive to the
atmospheric octant.
The discriminating observables between the two NO branches are therefore
$\theta_{23}$ and $\delta$ themselves, not $J_{\rm CP}$ or $m_{\beta\beta}$.
\begin{table}[htbp]
\centering
\caption{Branch-averaged predictions from the coefficient scan at
$\delta_\nu=270^\circ$ (mu--tau reflection); the last two rows add the
reflection-breaking band $\delta_\nu=270^\circ+\kappa$, $\kappa$ uniform on
$[-16^\circ,16^\circ]$ (Sec.~V~B).
We classify points by the atmospheric-octant branch (LO: $\theta_{23}<45^\circ$, HO: $\theta_{23}>45^\circ$).
We apply NuFIT-style cuts on $\sin^2\theta_{12}$ and $\sin^2\theta_{13}$ at the indicated level; the ``$2\sigma$'' band is implemented as a Gaussian proxy (twice the quoted $1\sigma$ widths), since the NuFIT $\delta$ and $\theta_{23}$ profiles are non-Gaussian for NO.
The effective Majorana mass $m_{\beta\beta}$ uses the $B$-lattice spectrum
($m_1=\epsilon^2 m_3\simeq 1.8~\mathrm{meV}$) with Majorana phases set to zero;
the reflection-symmetric values $(\alpha_{21},\alpha_{31})\in\{0,\pi\}$ give the
discrete set of Sec.~\ref{sec:0nubb}.
Entries are mean $\pm$ standard deviation.
Units: $\delta$ and $\theta_{23}$ are in degrees; $m_{\beta\beta}$ is in meV.}
\label{tab:branch-averages}
\scriptsize
\setlength{\tabcolsep}{2.8pt}
\renewcommand{\arraystretch}{1.05}
\resizebox{\columnwidth}{!}{
\begin{tabular}{lcccc}
Cut \& Branch & $\delta$ & $\theta_{23}$ & $m_{\beta\beta}$ & $J_{\rm CP}$ \\ \hline
$1\sigma$ LO ($N=7102$) & $259\pm7$ & $43.7\pm0.8$ & $2.8\pm0.1$ & $-0.0326\pm0.0008$ \\
$1\sigma$ HO ($N=4505$) & $272\pm9$ & $46.2\pm0.8$ & $2.8\pm0.1$ & $-0.0332\pm0.0006$ \\
\hline
$2\sigma$ LO ($N=25595$) & $260\pm7$ & $43.7\pm0.8$ & $2.8\pm0.2$ & $-0.0326\pm0.0010$ \\
$2\sigma$ HO ($N=18975$) & $274\pm10$ & $46.3\pm0.8$ & $2.9\pm0.1$ & $-0.0333\pm0.0008$ \\
\hline
$2\sigma$ LO, breaking band ($N=25438$) & $261\pm12$ & $43.7\pm0.8$ & $2.8\pm0.2$ & $-0.0321\pm0.0016$ \\
$2\sigma$ HO, breaking band ($N=18969$) & $274\pm14$ & $46.3\pm0.8$ & $2.9\pm0.2$ & $-0.0328\pm0.0013$ \\
\end{tabular}
}
\normalsize
\end{table}
\subsection{Analytic correlation formulas: \texorpdfstring{$\delta$--$\theta_{23}$}{delta--theta23} linkage}

\paragraph*{Leading-order octant--$\delta$ theorem.}
To leading nontrivial order in the small charged-lepton angles, the two-branch
structure in the $(\theta_{23},\delta)$ plane follows from a single complex
interference parameter controlling $U_{e3}$.
The full reactor element receives contributions from both the $12$ and $13$
charged-lepton rotations,
\begin{equation}
U_{e3} \simeq s_{13}^\nu e^{-i\delta_\nu}
- \theta_{12}^e\,s_{23}^\nu\,e^{i\phi_{12}^e}
- \theta_{13}^e\,c_{23}^\nu\,e^{i\phi_{13}^e}.
\label{eq:Ue3-full}
\end{equation}
In the $B$-lattice textures the ratio of the two correction terms is
$\theta_{13}^e c_{23}^\nu / (\theta_{12}^e s_{23}^\nu)
= \e^{1/3}\approx 0.57$, so the second correction is
comparable in size, not parametrically small.
Nevertheless, the dominant correlation structure is captured by
retaining only the larger $\theta_{12}^e$ term.
Define
\begin{equation}
r \equiv \frac{\theta_{12}^e\,s_{23}^\nu}{s_{13}^\nu},
\qquad
\Phi_e \equiv \phi_{12}^e - \delta_\nu,
\end{equation}
so that the reactor element reduces to
\begin{equation}
U_{e3} \simeq s_{13}^\nu e^{-i\delta_\nu}
\left(1 - r\,e^{-i\Phi_e}\right).
\end{equation}
Writing $U_{e3} = s_{13} e^{-i\delta}$ immediately yields
\begin{equation}
\boxed{
\delta \simeq \delta_\nu - \arg\!\left(1 - r\,e^{-i\Phi_e}\right)
}
\label{eq:delta-theorem}
\end{equation}
The observed Dirac phase $\delta$ is thus the neutrino-sector
phase $\delta_\nu$ shifted by a correction
$\arg(1-r\,e^{-i\Phi_e})$ that encodes the interference between the
neutrino-dominant reactor amplitude $s_{13}^\nu e^{-i\delta_\nu}$
and the charged-lepton correction
$\theta_{12}^e s_{23}^\nu e^{i\phi_{12}^e}$.
The magnitude and sign of this shift depend on two quantities:
the ratio $r$, which measures the relative size of the two
interfering amplitudes, and the effective phase difference
$\Phi_e=\phi_{12}^e-\delta_\nu$, which controls the direction
of the shift in the complex plane.
For the $B$-lattice value $r\sim\e^{8/9}\approx0.2$ at $c^e_{12}=1$, the
shift from this term is bounded by $\arcsin r\approx12^\circ$, displacing
$\delta$ from $\delta_\nu=270^\circ$ into the range
$\sim\!255^\circ$--$285^\circ$; over the full coefficient range, and with
the $\theta^e_{13}$ term included, the scan of Table~\ref{tab:branch-averages}
spans $240^\circ$--$296^\circ$ with dispersion $\sigma_\delta\simeq7$--$10^\circ$.

The atmospheric angle receives its dominant correction from the charged-lepton
$23$ rotation,
\begin{equation}
\boxed{\theta_{23} \simeq \theta_{23}^\nu - \theta_{23}^e \cos\phi_{23}^e.}
\label{eq:theta23-theorem}
\end{equation}
Thus the sign of $(\theta_{23}-\theta_{23}^\nu)$ is controlled by the same
charged-lepton phase that enters $U_{e3}$.

\paragraph*{Correlated branch structure.}
As shown in Sec.~\ref{sec:phase-alignment}, the single-flavon
origin of the $B$-lattice textures ensures
$\phi_{12}^e\approx\phi_{23}^e$ to within $\sim\!\phi_0/9$.
Under this alignment, the sign of the octant shift
$(\theta_{23}-45^\circ)$ correlates directly with the direction of the
phase displacement $(\delta-\delta_\nu)$: to first order
$\delta-\delta_\nu\simeq-r\sin(\phi^e-\delta_\nu)$ and
$\theta_{23}-45^\circ\simeq-\theta^e_{23}\cos\phi^e$ with
$\phi^e\equiv\phi^e_{12}\approx\phi^e_{23}$, and for $\delta_\nu=270^\circ$
the first becomes $-r\cos\phi^e$, so the two signs agree for every
$\phi^e$.
This correlation takes the compact form
\begin{equation}
\boxed{\;
\begin{aligned}
\operatorname{sgn}\!\left(\theta_{23}-45^\circ\right)
&=\operatorname{sgn}\!\left(\delta-\delta_\nu\right)\\[2pt]
&=\begin{cases}
-1, & \text{NO-I\ \ (lower quadrant)},\\[2pt]
+1, & \text{NO-II\ (upper quadrant)}.
\end{cases}
\end{aligned}
\;}
\label{eq:octant-quad}
\end{equation}
The two normal-ordering branches therefore map onto the lower and upper
quadrants of the $(\theta_{23}-45^\circ,\,\delta-\delta_\nu)$ plane,
corresponding to opposite signs of the interference term in
Eq.~\eqref{eq:delta-theorem} and producing the two branches in the
$(\theta_{23},\delta)$ plane.
The relation is exact in the aligned single-term limit at
$\delta_\nu=270^\circ$; in the full scan, with the $20^\circ$ misalignment
and the uniform $\phi^e_{13}$, it holds for $80\%$ of the points, and for
$98\%$ when the three charged-lepton phases are tied to $\phi_0$
(Table~\ref{tab:phase-variants}).  The agreement is a consequence of the
maximal neutrino phase: for $\delta_\nu=270^\circ+\Delta$ the signs disagree
over a fraction $|\Delta|/180^\circ$ of the phase circle, and the
correlation disappears at $\delta_\nu=0^\circ$ or $180^\circ$.

\paragraph*{Regime of validity.}
Equations~\eqref{eq:delta-theorem} and \eqref{eq:theta23-theorem} retain the
dominant charged-lepton corrections but neglect the
$\theta_{13}^e\,c_{23}^\nu$ term in Eq.~\eqref{eq:Ue3-full},
which is $\sim\!\e^{1/3}\approx 57\%$ of the retained
$\theta_{12}^e\,s_{23}^\nu$ term.
The analytic formulas should therefore be understood as capturing the
\emph{qualitative correlation structure} (the existence of two branches
and the sign of the octant--$\delta$ linkage) rather than as
precision predictions of $\delta(\theta_{23})$.
Numerical scans, which include all three charged-lepton rotations and their
relative phases, confirm that the two-branch structure and the predictive
octant--$\delta$ linkage are robust (Figs.~\ref{fig:scan-delta-theta23}
and \ref{fig:theta23-hist}).

The essential result is structural: the same interference that fixes
$|U_{e3}|$ determines both the magnitude and the direction of the Dirac-phase
shift and simultaneously selects the atmospheric octant.

\sbox{\figanglebox}{%
\begin{tikzpicture}
\draw[gray!18,line width=0.4pt] (0.000,4.750)--(5.000,4.750);
\draw[gray!18,line width=0.4pt] (0.000,5.327)--(5.000,5.327);
\draw[gray!18,line width=0.4pt] (0.000,5.904)--(5.000,5.904);
\draw[gray!18,line width=0.4pt] (0.000,6.481)--(5.000,6.481);
\draw[gray!18,line width=0.4pt] (0.000,7.058)--(5.000,7.058);
\draw[gray!18,line width=0.4pt] (0.000,7.635)--(5.000,7.635);
\begin{scope}
\clip (0.000,4.750) rectangle (5.000,7.750);
\fill[green!55!black!70] (0.329,4.750) rectangle (0.646,6.676);
\fill[green!55!black!70] (1.995,4.750) rectangle (2.313,7.230);
\fill[green!55!black!70] (3.662,4.750) rectangle (3.979,5.245);
\fill[violet!65] (0.675,4.750) rectangle (0.992,6.676);
\fill[violet!65] (2.341,4.750) rectangle (2.659,7.416);
\fill[violet!65] (4.008,4.750) rectangle (4.325,5.245);
\fill[blue!22] (1.021,4.750) rectangle (1.338,6.712);
\fill[blue!22] (2.687,4.750) rectangle (3.005,7.346);
\fill[blue!22] (4.354,4.750) rectangle (4.671,5.272);
\draw[dashed,gray,line width=0.6pt] (0.100,6.784)--(1.567,6.784);
\draw[dashed,gray,line width=0.6pt] (1.767,7.346)--(3.233,7.346);
\end{scope}
\draw[line width=0.6pt] (0.000,4.750) rectangle (5.000,7.750);
\draw[line width=0.5pt] (0.000,4.750)--(-0.070,4.750); \node[left,font=\scriptsize] at (-0.090,4.750) {0};
\draw[line width=0.5pt] (0.000,5.327)--(-0.070,5.327); \node[left,font=\scriptsize] at (-0.090,5.327) {10};
\draw[line width=0.5pt] (0.000,5.904)--(-0.070,5.904); \node[left,font=\scriptsize] at (-0.090,5.904) {20};
\draw[line width=0.5pt] (0.000,6.481)--(-0.070,6.481); \node[left,font=\scriptsize] at (-0.090,6.481) {30};
\draw[line width=0.5pt] (0.000,7.058)--(-0.070,7.058); \node[left,font=\scriptsize] at (-0.090,7.058) {40};
\draw[line width=0.5pt] (0.000,7.635)--(-0.070,7.635); \node[left,font=\scriptsize] at (-0.090,7.635) {50};
\draw[line width=0.5pt] (0.833,4.750)--(0.833,4.680); \node[below,font=\scriptsize] at (0.833,4.660) {$\theta_{12}$};
\draw[line width=0.5pt] (2.500,4.750)--(2.500,4.680); \node[below,font=\scriptsize] at (2.500,4.660) {$\theta_{23}$};
\draw[line width=0.5pt] (4.167,4.750)--(4.167,4.680); \node[below,font=\scriptsize] at (4.167,4.660) {$\theta_{13}$};
\node[font=\scriptsize,rotate=90] at (-0.620,6.250) {deg};
\node[anchor=north west,font=\footnotesize,fill=white,inner sep=1pt] at (0.080,7.670) {(a)};
\fill[green!55!black!70] (0.150,7.960) rectangle (0.310,8.100); \node[right,font=\scriptsize] at (0.350,8.030) {NO-I};
\fill[violet!65] (1.817,7.960) rectangle (1.977,8.100); \node[right,font=\scriptsize] at (2.017,8.030) {NO-II};
\fill[blue!22] (3.483,7.960) rectangle (3.643,8.100); \node[right,font=\scriptsize] at (3.683,8.030) {$\nu$};
\draw[gray!18,line width=0.4pt] (6.950,4.750)--(11.950,4.750);
\draw[gray!18,line width=0.4pt] (6.950,5.734)--(11.950,5.734);
\draw[gray!18,line width=0.4pt] (6.950,6.717)--(11.950,6.717);
\draw[gray!18,line width=0.4pt] (6.950,7.701)--(11.950,7.701);
\begin{scope}
\clip (6.950,4.750) rectangle (11.950,7.750);
\fill[blue!22] (7.348,4.750) rectangle (7.765,7.116);
\fill[blue!22] (9.015,4.750) rectangle (9.431,6.714);
\fill[blue!22] (10.681,4.750) rectangle (11.098,6.102);
\fill[green!55!black!70] (7.802,4.750) rectangle (8.219,6.643);
\fill[green!55!black!70] (9.469,4.750) rectangle (9.885,6.871);
\fill[green!55!black!70] (11.135,4.750) rectangle (11.552,5.832);
\end{scope}
\draw[line width=0.6pt] (6.950,4.750) rectangle (11.950,7.750);
\draw[line width=0.5pt] (6.950,4.750)--(6.880,4.750); \node[left,font=\scriptsize] at (6.860,4.750) {0};
\draw[line width=0.5pt] (6.950,5.734)--(6.880,5.734); \node[left,font=\scriptsize] at (6.860,5.734) {1};
\draw[line width=0.5pt] (6.950,6.717)--(6.880,6.717); \node[left,font=\scriptsize] at (6.860,6.717) {2};
\draw[line width=0.5pt] (6.950,7.701)--(6.880,7.701); \node[left,font=\scriptsize] at (6.860,7.701) {3};
\draw[line width=0.5pt] (7.783,4.750)--(7.783,4.680); \node[below,font=\scriptsize] at (7.783,4.660) {$\theta^e_{12}$};
\draw[line width=0.5pt] (9.450,4.750)--(9.450,4.680); \node[below,font=\scriptsize] at (9.450,4.660) {$\theta^e_{23}$};
\draw[line width=0.5pt] (11.117,4.750)--(11.117,4.680); \node[below,font=\scriptsize] at (11.117,4.660) {$\theta^e_{13}$};
\node[font=\scriptsize,rotate=90] at (6.330,6.250) {deg};
\node[anchor=north west,font=\footnotesize,fill=white,inner sep=1pt] at (7.030,7.670) {(b)};
\fill[blue!22] (7.100,7.960) rectangle (7.260,8.100); \node[right,font=\scriptsize] at (7.300,8.030) {base};
\fill[green!55!black!70] (9.600,7.960) rectangle (9.760,8.100); \node[right,font=\scriptsize] at (9.800,8.030) {bench};
\node[font=\tiny] at (7.783,7.327) {$\epsilon^{17/9}$};
\node[font=\tiny] at (9.450,7.327) {$\epsilon^{2}$};
\node[font=\tiny] at (11.117,7.327) {$\epsilon^{20/9}$};
\draw[gray!18,line width=0.4pt] (0.000,0.000)--(5.000,0.000);
\draw[gray!18,line width=0.4pt] (0.000,0.577)--(5.000,0.577);
\draw[gray!18,line width=0.4pt] (0.000,1.154)--(5.000,1.154);
\draw[gray!18,line width=0.4pt] (0.000,1.731)--(5.000,1.731);
\draw[gray!18,line width=0.4pt] (0.000,2.308)--(5.000,2.308);
\draw[gray!18,line width=0.4pt] (0.000,2.885)--(5.000,2.885);
\begin{scope}
\clip (0.000,0.000) rectangle (5.000,3.000);
\fill[green!55!black!70] (0.398,0.000) rectangle (0.815,0.721);
\fill[green!55!black!70] (2.065,0.000) rectangle (2.481,2.519);
\fill[green!55!black!70] (3.731,0.000) rectangle (4.148,2.093);
\fill[violet!65] (0.852,0.000) rectangle (1.269,0.682);
\fill[violet!65] (2.519,0.000) rectangle (2.935,2.702);
\fill[violet!65] (4.185,0.000) rectangle (4.602,1.910);
\draw[dashed,gray,line width=0.6pt] (1.767,2.596)--(3.233,2.596);
\draw[dashed,gray,line width=0.6pt] (3.433,2.034)--(4.900,2.034);
\end{scope}
\draw[line width=0.6pt] (0.000,0.000) rectangle (5.000,3.000);
\draw[line width=0.5pt] (0.000,0.000)--(-0.070,0.000); \node[left,font=\scriptsize] at (-0.090,0.000) {0};
\draw[line width=0.5pt] (0.000,0.577)--(-0.070,0.577); \node[left,font=\scriptsize] at (-0.090,0.577) {10};
\draw[line width=0.5pt] (0.000,1.154)--(-0.070,1.154); \node[left,font=\scriptsize] at (-0.090,1.154) {20};
\draw[line width=0.5pt] (0.000,1.731)--(-0.070,1.731); \node[left,font=\scriptsize] at (-0.090,1.731) {30};
\draw[line width=0.5pt] (0.000,2.308)--(-0.070,2.308); \node[left,font=\scriptsize] at (-0.090,2.308) {40};
\draw[line width=0.5pt] (0.000,2.885)--(-0.070,2.885); \node[left,font=\scriptsize] at (-0.090,2.885) {50};
\draw[line width=0.5pt] (0.833,0.000)--(0.833,-0.070); \node[below,font=\scriptsize] at (0.833,-0.090) {$\theta_\ell$};
\draw[line width=0.5pt] (2.500,0.000)--(2.500,-0.070); \node[below,font=\scriptsize] at (2.500,-0.090) {$\theta$};
\draw[line width=0.5pt] (4.167,0.000)--(4.167,-0.070); \node[below,font=\scriptsize] at (4.167,-0.090) {$\theta_\nu$};
\node[font=\scriptsize,rotate=90] at (-0.620,1.500) {deg};
\node[anchor=north west,font=\footnotesize,fill=white,inner sep=1pt] at (0.080,2.920) {(c)};
\fill[green!55!black!70] (0.150,3.210) rectangle (0.310,3.350); \node[right,font=\scriptsize] at (0.350,3.280) {NO-I};
\fill[violet!65] (2.650,3.210) rectangle (2.810,3.350); \node[right,font=\scriptsize] at (2.850,3.280) {NO-II};
\draw[gray!18,line width=0.4pt] (6.950,0.000)--(11.950,0.000);
\draw[gray!18,line width=0.4pt] (6.950,0.698)--(11.950,0.698);
\draw[gray!18,line width=0.4pt] (6.950,1.395)--(11.950,1.395);
\draw[gray!18,line width=0.4pt] (6.950,2.093)--(11.950,2.093);
\draw[gray!18,line width=0.4pt] (6.950,2.791)--(11.950,2.791);
\begin{scope}
\clip (6.950,0.000) rectangle (11.950,3.000);
\fill[green!55!black!70] (7.348,0.698) rectangle (7.765,2.477);
\fill[green!55!black!70] (9.015,0.698) rectangle (9.431,0.220);
\fill[green!55!black!70] (10.681,0.698) rectangle (11.098,2.581);
\fill[violet!65] (7.802,0.698) rectangle (8.219,2.621);
\fill[violet!65] (9.469,0.698) rectangle (9.885,0.080);
\fill[violet!65] (11.135,0.698) rectangle (11.552,2.581);
\draw[dashed,gray,line width=0.6pt] (8.717,0.070)--(10.183,0.070);
\draw[dashed,gray,line width=0.6pt] (10.383,2.581)--(11.850,2.581);
\draw[dashed,gray,line width=0.6pt] (7.050,2.581)--(8.517,2.581);
\end{scope}
\draw[line width=0.6pt] (6.950,0.000) rectangle (11.950,3.000);
\draw[line width=0.5pt] (6.950,0.000)--(6.880,0.000); \node[left,font=\scriptsize] at (6.860,0.000) {-100};
\draw[line width=0.5pt] (6.950,0.698)--(6.880,0.698); \node[left,font=\scriptsize] at (6.860,0.698) {0};
\draw[line width=0.5pt] (6.950,1.395)--(6.880,1.395); \node[left,font=\scriptsize] at (6.860,1.395) {100};
\draw[line width=0.5pt] (6.950,2.093)--(6.880,2.093); \node[left,font=\scriptsize] at (6.860,2.093) {200};
\draw[line width=0.5pt] (6.950,2.791)--(6.880,2.791); \node[left,font=\scriptsize] at (6.860,2.791) {300};
\draw[line width=0.5pt] (7.783,0.000)--(7.783,-0.070); \node[below,font=\scriptsize] at (7.783,-0.090) {$\delta$};
\draw[line width=0.5pt] (9.450,0.000)--(9.450,-0.070); \node[below,font=\scriptsize] at (9.450,-0.090) {$\phi_{\rm FX}$};
\draw[line width=0.5pt] (11.117,0.000)--(11.117,-0.070); \node[below,font=\scriptsize] at (11.117,-0.090) {$\delta_\nu$};
\node[font=\scriptsize,rotate=90] at (6.050,1.500) {deg};
\node[anchor=north west,font=\footnotesize,fill=white,inner sep=1pt] at (7.030,2.920) {(d)};
\fill[green!55!black!70] (7.100,3.210) rectangle (7.260,3.350); \node[right,font=\scriptsize] at (7.300,3.280) {NO-I};
\fill[violet!65] (9.600,3.210) rectangle (9.760,3.350); \node[right,font=\scriptsize] at (9.800,3.280) {NO-II};
\end{tikzpicture}%
}
\begin{figure*}[htbp]
\centering
\resizebox{\textwidth}{!}{%
\usebox{\figanglebox}%
}
\caption{Summary of mixing angles and phases in the lattice-flavon
lepton model.
\emph{Top row:} PDG mixing angles $\theta_{12}$, $\theta_{23}$,
$\theta_{13}$ compared with the neutrino-sector values
($\theta_{12}^\nu$, $\theta_{23}^\nu$, $\theta_{13}^\nu$)
for the two normal-ordering branches (NO-I and NO-II);
dashed lines mark the tribimaximal and maximal reference values; $\theta_{12}$ falls just below tribimaximal and $\theta_{23}$ straddles maximal mixing.
\emph{Middle row:} charged-lepton correction angles
$\theta_{ij}^e$ at the $B$-lattice base scale ($c=1$, blue bars)
and at the benchmark values (green bars), on an expanded vertical
scale; labels indicate the $\epsilon$-power scaling.
\emph{Bottom left:} Fritzsch--Xing rotation angles
$(\theta_\ell,\,\theta,\,\theta_\nu)$ for NO-I (green) and NO-II
(purple), with tribimaximal and maximal reference lines; $\theta$ straddles maximal mixing and $\theta_\nu$ straddles the tribimaximal value ($+1.0^\circ$ for NO-I, $-2.2^\circ$ for NO-II).
\emph{Bottom right:} Dirac phase $\delta$, FX phase $\phi_{\rm FX}$,
and neutrino-sector phase $\delta_\nu$ for both branches;
the dashed lines mark the quark-sector reference $\pm 90^\circ$ and $270^\circ$; $\delta_\nu=270^\circ$ coincides with the reference, $\delta$ lies $15^\circ$ below it (NO-I) and $6^\circ$ above it (NO-II), and $\phi_{\rm FX}$ is within $2^\circ$ of $-90^\circ$ for NO-II and $21^\circ$ above it for NO-I.}
\label{fig:angle-summary}
\end{figure*}

Figure~\ref{fig:angle-summary} collects the full set of mixing angles
and CP phases, providing a compact visual comparison of the two
normal-ordering branches and the hierarchy of charged-lepton corrections.

\subsection{Geometric form: the atmospheric right triangle}
\label{sec:atm-triangle}

The octant selection has a compact geometric statement. The third-column
normalization $|U_{e3}|^2+|U_{\mu3}|^2+|U_{\tau3}|^2=1$ fixes the
$\mu$--$\tau$ block $(U_{\mu3},U_{\tau3})=c_{13}(s_{23},c_{23})$, from which one
forms the right triangle of Fig.~\ref{fig:atm-triangle} with legs
\begin{equation}
\begin{aligned}
|U_{\tau3}|^2-|U_{\mu3}|^2 &= c_{13}^2\cos2\theta_{23},\\
2\,|U_{\mu3}|\,|U_{\tau3}| &= c_{13}^2\sin2\theta_{23},
\end{aligned}
\label{eq:atm-legs}
\end{equation}
hypotenuse $|U_{\mu3}|^2+|U_{\tau3}|^2=c_{13}^2$, and angle $2\theta_{23}$
between the hypotenuse and the first leg. All three side lengths are
determined by the PMNS parameters through Eq.~\eqref{eq:atm-legs} and close
into a triangle, the closure being the Pythagorean identity
$\cos^2 2\theta_{23}+\sin^2 2\theta_{23}=1$; it is a $\nu_3$-column
normalization triangle, its sides fixed by the third-column normalization
rather than by the orthogonality of two columns of $U_{\rm PMNS}$ that defines
the standard unitarity triangles. The right angle, where the first leg
vanishes, is the maximal-mixing point $\theta_{23}=45^\circ$
($2\theta_{23}=90^\circ$), at which $|U_{\mu3}|=|U_{\tau3}|$, the
$\mu$--$\tau$ symmetry of the third column. The octant is the sign of that
first leg, the $\mu$--$\tau$ asymmetry $|U_{\tau3}|^2-|U_{\mu3}|^2$, positive
for NO-I ($2\theta_{23}\approx86^\circ$, lower octant) and negative for NO-II
($2\theta_{23}\approx92^\circ$, upper octant); the apex crosses
maximal mixing as the branch changes, giving the two atmospheric right
triangles, one in each octant.

The first leg is the charged-lepton octant correction. With
$\theta_{23}-45^\circ=-\theta_{23}^e\cos\phi_{23}^e$ from
Eq.~\eqref{eq:theta23-theorem}, the small-deviation expansion gives
$\cos2\theta_{23}\simeq2\theta_{23}^e\cos\phi_{23}^e$, so
\begin{equation}
|U_{\tau3}|^2-|U_{\mu3}|^2\simeq 2\,c_{13}^2\,\theta_{23}^e\cos\phi_{23}^e,
\label{eq:atm-leg-cl}
\end{equation}
with sign $\operatorname{sgn}(\cos\phi_{23}^e)$. The same charged-lepton phase
that selects the octant in Eq.~\eqref{eq:octant-quad} orients the apex, so
Fig.~\ref{fig:atm-triangle} is the geometric form of the octant--$\delta$ box,
with the right angle as the symmetric reference, the horizontal leg as the
charged-lepton correction, and the branch as the side of the right angle on
which $2\theta_{23}$ falls.

\begin{figure}[htbp]
\centering
\begin{tikzpicture}[font=\small,>=Latex]
  \def\R{3.4}
  \coordinate (O) at (0,0);
  \coordinate (P) at (78:\R);   \coordinate (F)  at (78:\R |- 0,0);
  \coordinate (Pp) at (94.6:\R); \coordinate (Fp) at (94.6:\R |- 0,0);
  \draw[gray!45,line width=0.5pt] (-1.5,0) -- (1.6,0);
  \node[above,align=center] at (0,\R+1.1) {near $\mu$--$\tau$ maximal mixing\\[1pt]$\theta_{23}\approx45^\circ$ and $2\theta_{23}\approx90^\circ$};
  \draw[gray!50] (70:\R) arc (70:104:\R);
  \draw[dashed,line width=0.8pt] (O) -- (Pp) -- (Fp) -- cycle;
  \fill (Pp) circle (1.3pt); \node[above left=-1pt] at (Pp) {upper octant (NO-II)};
  \draw[line width=1pt]
      (O) -- (P)
          -- (F)  node[sloped,above,pos=0.5]{$c_{13}^2\sin2\theta_{23}$}
          -- (O)  node[sloped,below,pos=0.5,yshift=-1pt]{$c_{13}^2\cos2\theta_{23}$} -- cycle;
  \fill (P) circle (1.3pt);
  \node at (0.22,\R+0.25) {$c_{13}^2$};
  \node[right=1pt] at (P) {lower octant (NO-I)};
\end{tikzpicture}
\caption{The atmospheric right triangle, from the third-column normalization
$|U_{e3}|^2+|U_{\mu3}|^2+|U_{\tau3}|^2=1$. The $\mu$--$\tau$ block of $\nu_3$
forms a right triangle with adjacent leg
$|U_{\tau3}|^2-|U_{\mu3}|^2=c_{13}^2\cos2\theta_{23}$ (the $\mu$--$\tau$
asymmetry), opposite leg $2|U_{\mu3}||U_{\tau3}|=c_{13}^2\sin2\theta_{23}$,
hypotenuse $|U_{\mu3}|^2+|U_{\tau3}|^2=c_{13}^2$, and angle $2\theta_{23}$ at the shared lower vertex where each hypotenuse meets the asymmetry axis.
This angle reaches $90^\circ$ at maximal mixing,
$\theta_{23}=45^\circ$, where the asymmetry leg vanishes. The octant is the sign of the adjacent leg, with NO-I
(lower octant) placing the apex to the right of maximal mixing and NO-II
(upper octant) to its left. The departures from maximal mixing are exaggerated
for clarity by a common factor, so their relative size is preserved; NO-I
lies farther below maximal than NO-II lies above. This is the geometric form of
Eq.~\eqref{eq:octant-quad}.}
\label{fig:atm-triangle}
\end{figure}

\section{Parameterization by increments about the tribimaximal limit}
\label{sec:increments}

The two normal-ordering branches are reached from a single reference point at
which the lepton mixing is exactly tribimaximal and the Dirac phase maximal,
\begin{equation}
\begin{aligned}
(\theta_\ell,\,\theta,\,\theta_\nu,\,\delta)_{\rm ref}
&=\bigl(0,\;45^\circ,\;\theta_\nu^{\rm TBM},\;270^\circ\bigr),\\
\theta_\nu^{\rm TBM}&=\arctan(1/\!\sqrt2)\simeq 35.3^\circ,
\end{aligned}
\label{eq:ref-point}
\end{equation}
for which $\theta_{13}=0$, $\theta_{23}=45^\circ$, $\sin^2\theta_{12}=\tfrac13$,
and the leptonic CP violation is maximal. Here $\theta_\ell$, $\theta$, and
$\theta_\nu$ are the Fritzsch--Xing rotation angles~\cite{Fritzsch} and $\delta$
is the physical Dirac phase; the measured mixing follows by adding the
increments of Table~\ref{tab:increments}.

\begin{table}[t]
\centering
\caption{Increments of the mixing parameters relative to the tribimaximal,
maximal-CP reference of Eq.~\eqref{eq:ref-point}.}
\label{tab:increments}
\begin{tabular}{lcc}
\hline\hline
Increment & NO-I & NO-II \\ \hline
$\theta_\ell$ & $+12.5^\circ$ & $+11.8^\circ$ \\
$\theta-45^\circ$ & $-1.3^\circ$ & $+1.8^\circ$ \\
$\theta_\nu-\theta_\nu^{\rm TBM}$ & $+1.0^\circ$ & $-2.2^\circ$ \\
$\delta-270^\circ$ & $-15^\circ$ & $+6^\circ$ \\
\hline\hline
\end{tabular}
\end{table}

The four increments do not share a common origin; they divide between the two
sectors. Two are neutrino-sector effects of the $\mathcal{O}(\e)$ first row of
the reflection-symmetric texture. The reactor angle rests on the neutrino amplitude $s_{13}^\nu\sim\e$; with
$\theta_{13}^\nu\approx0.15$ ($\sim\!\e$, from $\mathrm{Im}\,r$ in
Eq.~\eqref{eq:mnu-LO}) it gives $\theta_{13}\approx8.6^\circ$, reproducing the observed
$8.5^\circ$ with a coefficient near unity. The charged-lepton $12$-rotation is
steeply suppressed, $\theta_{12}^e\sim\e^{17/9}\approx2.5^\circ$, smaller than the
neutrino amplitude by $\sim\!\e^{8/9}\approx0.2$, so it enters the reactor element
$U_{e3}\simeq s_{13}^\nu e^{-i\delta_\nu}(1-r\,e^{-i\Phi_e})$ only as the
interference $r=\theta_{12}^e s_{23}^\nu/s_{13}^\nu\sim\e^{8/9}$ that sets the
modest Dirac-phase shift [Eq.~\eqref{eq:delta-theorem}]. The effective FX angle
$\theta_\ell\approx12^\circ$ of the identity
$\theta_{13}=\sin\theta_\ell\sin\theta$ is therefore $\sqrt2\,\theta_{13}$, fixed
by this neutrino-dominant reactor angle and not by the physical charged-lepton
angle $\theta_{12}^e$; its apparent size reflects the factorization, not a large
charged-lepton rotation. The Dirac phase is
likewise neutrino-led, $\delta\simeq\delta_\nu-\arg(1-r\,e^{-i\Phi_e})$ with
$\delta_\nu=270^\circ$ fixed by reflection, so the displacement from
$270^\circ$ is entirely the charged-lepton interference, $-15^\circ$ (NO-I)
and $+6^\circ$ (NO-II). The other two increments are
charged-lepton corrections to a maximal, tribimaximal neutrino starting point.
The octant is the sign of $\theta-45^\circ\simeq-\theta_{23}^e\cos\phi_{23}^e$
[Eq.~\eqref{eq:theta23-theorem}], $1.3^\circ$ below (NO-I) and $1.8^\circ$
above (NO-II) maximal, and the FX solar angle, with
$\theta_{12}^\nu\approx\theta_\nu^{\rm TBM}$, is raised $1.0^\circ$ (NO-I)
or lowered $2.2^\circ$ (NO-II) by the charged-lepton $12$-rotation
together with the FX repackaging that accompanies $\theta_{13}$. The two-branch
structure is then a property of the correlated sign of the charged-lepton
interference, which ties the octant of $\theta_{23}$ to the side of $\delta_\nu$
on which $\delta$ falls [Eq.~\eqref{eq:octant-quad}], not of any single
increment.

\section{Conclusions}
\label{sec:conclusions}

We have extended the single-flavon $B$-lattice framework to the lepton sector. The construction rests on two ingredients: (i)~$B$-lattice power counting ($\epsilon=1/B$), which fixes charged-lepton Yukawa suppressions and neutrino mass eigenvalue ratios, and (ii)~mu--tau reflection symmetry of the neutrino mass matrix~\cite{HarrisonScott,GrimusLavoura,MuTauReview}, a generalized CP symmetry realizable within $S_4$ with CP~\cite{FHZ,HLS,A4Review,IshimoriReview}, which fixes the $\mathcal{O}(1)$ structure of the $23$ block, the maximal atmospheric angle, and the maximal neutrino phase $\delta_\nu=270^\circ$, while leaving the reactor angle free at $\mathcal{O}(\epsilon)$. The charged-lepton sector requires ingredient~(i) together with the off-diagonal suppression of Appendix~\ref{app:charges-scan}, which a generation-dependent selection rule supplies; the neutrino sector requires both (i) and (ii).

The central result is a two-branch prediction in the $(\theta_{23},\delta)$ plane, the two lobes of one continuous distribution:
\begin{itemize}
\item \textbf{NO-I} (lower octant): $\theta_{23}\approx 44^\circ$, $\delta\approx 260^\circ$, mildly favored in the coefficient scan and containing the strict single-source point $\phi_0=\pi$;
\item \textbf{NO-II} (upper octant): $\theta_{23}\approx 46^\circ$, $\delta\approx 275^\circ$.
\end{itemize}
The two branches arise because the same charged-lepton--neutrino interference that generates the reactor element $U_{e3}$ also controls the direction of the Dirac-phase shift $\delta\simeq\delta_\nu-\arg(1-r\,e^{-i\Phi_e})$ [Eq.~\eqref{eq:delta-theorem}]; the single-flavon origin of the Yukawa textures ensures the required phase alignment between the $12$ and $23$ charged-lepton rotations to within $\sim\!\phi_0/9$. This correlation is stable under $\mathcal{O}(1)$ coefficient variations. Geometrically, the same octant--$\delta$ structure is a $\nu_3$-column normalization triangle (Fig.~\ref{fig:atm-triangle}), with the $\nu_3$ column forming a right triangle whose base angle is $2\theta_{23}$; maximal mixing is the $90^\circ$ base angle, and the octant is set by whether the base angle lies below or above $90^\circ$.

Three observables are nearly branch-independent and therefore cannot discriminate: $J_{\rm CP}\simeq -0.033$, $m_{\beta\beta}\simeq 3~\mathrm{meV}$ (for vanishing Majorana phases), and $\sin^2\theta_{13}\simeq 0.022$. The discriminating observables are $\theta_{23}$ and $\delta$ themselves.

The framework makes a falsifiable prediction for the Dirac phase. Mu--tau reflection fixes $\delta_\nu=270^\circ$ at leading order, so that $\delta$ is $270^\circ$ shifted by the charged-lepton interference, with the sign and size of the shift correlated with the octant: $\delta\approx260^\circ$ in the lower octant and $\approx275^\circ$ in the upper, and in all cases within about $40^\circ$ of $270^\circ$, i.e.\ $|J_{\rm CP}|$ at least $70\%$ of its maximal value. A Dirac phase in the first or second quadrant, CP conservation, a phase more than about $40^\circ$ from maximal, or $\sin^2\theta_{23}$ outside $0.44$--$0.56$ would each exclude the single-$B$ textures with reflection-symmetric neutrinos; a lower octant with $\delta>270^\circ$ or an upper octant with $\delta<270^\circ$ would exclude the aligned-phase charged-lepton structure. If the reflection symmetry were dropped and $\delta_\nu$ left free, the framework would still confine $\sin^2\theta_{23}$ to $0.44$--$0.56$ and $\delta$ to within about $25^\circ$ of $\delta_\nu$, but $\delta$ alone could not test it. The present normal-ordering global fit, consistent with CP conservation within $1\sigma$, is in mild tension with this prediction, while T2K favors its center; the first appearance data from DUNE and Hyper-Kamiokande will decide it. The decisive measurements are $\nu_\mu\to\nu_e$ appearance rates and their CP conjugates at DUNE~\cite{DUNE} and Hyper-Kamiokande~\cite{HyperK}, which are directly sensitive to $\sin\delta$ through the Jarlskog interference term, combined with octant determination from $\nu_\mu$ disappearance and atmospheric neutrino samples at IceCube~\cite{IceCube}. Independent resolution of the mass ordering by JUNO~\cite{JUNO} eliminates the $\delta\leftrightarrow(\pi-\delta)$ degeneracy from matter effects. Together these measurements, building on current constraints from Super-Kamiokande~\cite{SuperK}, NOvA~\cite{NOvA}, and T2K~\cite{T2K}, will provide a direct test of this organizing principle across quarks and leptons.
\appendix
\section{Analytic PMNS matrix in terms of model parameters}
\label{app:PMNS-analytic}

We collect here the explicit construction of the PMNS matrix
used throughout this paper.

\subsection{Model parameters}

The framework has two sectors, each specified by a small number of parameters.

\paragraph{Neutrino sector (4 parameters).}
The neutrino diagonalization matrix $\Unu$ is parameterized in
the PDG convention,
\begin{equation}
\Unu = R_{23}(\theta_{23}^\nu)\,
       R_{13}(\theta_{13}^\nu,\,\delta_\nu)\,
       R_{12}(\theta_{12}^\nu),
\label{eq:Unu-param}
\end{equation}
with $\theta_{23}^\nu=45^\circ$ and $\delta_\nu=270^\circ$ (exact, from mu--tau
reflection), $\theta_{12}^\nu\simeq 35^\circ$ (near-tribimaximal), and
$\theta_{13}^\nu\sim\e$ (numerically $\approx 0.15$~rad, from the imaginary
first-row entry of Eq.~\eqref{eq:mnu-LO}).

\paragraph{Charged-lepton sector (6 parameters).}
The charged-lepton diagonalization matrix $\Ue$ is built from
three small rotations with phases,
\begin{equation}
\Ue = R_{23}(\theta_{23}^e,\,\phi_{23}^e)\,
      R_{13}(\theta_{13}^e,\,\phi_{13}^e)\,
      R_{12}(\theta_{12}^e,\,\phi_{12}^e),
\label{eq:Ue-param}
\end{equation}
where the rotation angles are fixed by the $B$-lattice exponent
differences up to $\mathcal{O}(1)$ coefficients:
\begin{equation}
\theta_{12}^e = c_{12}^e\,\e^{17/9},\quad
\theta_{23}^e = c_{23}^e\,\e^{2},\quad
\theta_{13}^e = c_{13}^e\,\e^{20/9},
\label{eq:theta-e-scaling}
\end{equation}
with $c_{ij}^e=\mathcal{O}(1)$.
The phases $\phi_{ij}^e$ arise from the complex flavon VEV
as described in Sec.~\ref{sec:phase-alignment}; the single-flavon
counting gives the approximate alignment
$\phi_{12}^e\approx\phi_{23}^e$ with misalignment
$\sim\!\phi_0/9$, and the realization of Appendix~\ref{app:charges-scan}
makes it exact.

\paragraph{Numerical base values.}
With $\e=1/B\simeq 0.1867$:
\begin{equation}
\begin{aligned}
\theta_{12}^e\big|_{c=1} &= \e^{17/9} \simeq 0.042~\text{rad}\simeq 2.4^\circ,\\
\theta_{23}^e\big|_{c=1} &= \e^{2}\phantom{/9} \simeq 0.035~\text{rad}\simeq 2.0^\circ,\\
\theta_{13}^e\big|_{c=1} &= \e^{20/9} \simeq 0.024~\text{rad}\simeq 1.4^\circ.
\end{aligned}
\label{eq:base-angles}
\end{equation}

\subsection{Rotation matrix conventions}

Each rotation $R_{ij}(\theta,\phi)$ acts in the $ij$ plane with
a phase $\phi$ attached to the off-diagonal sine entries.
Explicitly,
\begin{equation}
R_{12}(\theta,\phi)=
\begin{pmatrix}
c & s\,e^{i\phi} & 0\\
-s\,e^{-i\phi} & c & 0\\
0 & 0 & 1
\end{pmatrix}\!,
\end{equation}
\begin{equation}
R_{13}(\theta,\phi)=
\begin{pmatrix}
c & 0 & s\,e^{i\phi}\\
0 & 1 & 0\\
-s\,e^{-i\phi} & 0 & c
\end{pmatrix}\!,
\end{equation}
\begin{equation}
R_{23}(\theta,\phi)=
\begin{pmatrix}
1 & 0 & 0\\
0 & c & s\,e^{i\phi}\\
0 & -s\,e^{-i\phi} & c
\end{pmatrix}\!,
\end{equation}
where $c\equiv\cos\theta$, $s\equiv\sin\theta$.
When the phase argument is omitted, $\phi=0$.

\subsection{Leading-order PMNS matrix}

Since all charged-lepton angles satisfy
$\theta_{ij}^e\ll 1$, $\Ue$ may be expanded to first order:
\begin{equation}
\Ue\simeq\mathbb{1}+
\begin{pmatrix}
0 & \theta_{12}^e\,e^{i\phi_{12}^e}
  & \theta_{13}^e\,e^{i\phi_{13}^e}\\[4pt]
-\theta_{12}^e\,e^{-i\phi_{12}^e} & 0
  & \theta_{23}^e\,e^{i\phi_{23}^e}\\[4pt]
-\theta_{13}^e\,e^{-i\phi_{13}^e}
  & -\theta_{23}^e\,e^{-i\phi_{23}^e} & 0
\end{pmatrix}
\!+\mathcal{O}(\e^{34/9}).
\label{eq:Ue-expand}
\end{equation}
The Hermitian conjugate is
\begin{equation}
\Ue^\dagger\simeq\mathbb{1}+
\begin{pmatrix}
0 & -\theta_{12}^e\,e^{i\phi_{12}^e}
  & -\theta_{13}^e\,e^{i\phi_{13}^e}\\[4pt]
\theta_{12}^e\,e^{-i\phi_{12}^e} & 0
  & -\theta_{23}^e\,e^{i\phi_{23}^e}\\[4pt]
\theta_{13}^e\,e^{-i\phi_{13}^e}
  & \theta_{23}^e\,e^{-i\phi_{23}^e} & 0
\end{pmatrix}\!.
\label{eq:Ued-expand}
\end{equation}

The PMNS matrix $\UPMNS=\Ue^\dagger\Unu$ then takes the form
\begin{equation}
(\UPMNS)_{\alpha i}
\;=\;(\Unu)_{\alpha i}
+ \sum_{\beta\neq\alpha}
(\Delta_e^\dagger)_{\alpha\beta}\,(\Unu)_{\beta i}
+\mathcal{O}(\e^{34/9}),
\label{eq:PMNS-expand}
\end{equation}
where $\Delta_e^\dagger\equiv\Ue^\dagger-\mathbb{1}$ is the
off-diagonal perturbation in Eq.~\eqref{eq:Ued-expand}.

\subsection{Explicit PMNS elements}

Using the shorthand $(\Unu)_{\alpha i}\equiv U^\nu_{\alpha i}$ and
suppressing the superscript $e$ on the charged-lepton angles and phases,
the nine elements are:
\begin{align}
U_{e1} &= U^\nu_{e1}
 - \theta_{12}\,e^{i\phi_{12}}\,U^\nu_{\mu 1}
 - \theta_{13}\,e^{i\phi_{13}}\,U^\nu_{\tau 1},
\label{eq:Ue1}\\[4pt]
U_{e2} &= U^\nu_{e2}
 - \theta_{12}\,e^{i\phi_{12}}\,U^\nu_{\mu 2}
 - \theta_{13}\,e^{i\phi_{13}}\,U^\nu_{\tau 2},
\label{eq:Ue2}\\[4pt]
U_{e3} &= U^\nu_{e3}
 - \theta_{12}\,e^{i\phi_{12}}\,U^\nu_{\mu 3}
 - \theta_{13}\,e^{i\phi_{13}}\,U^\nu_{\tau 3}
\nonumber\\
&= s_{13}^\nu\,e^{-i\delta_\nu}
 - \theta_{12}\,s_{23}^\nu c_{13}^\nu\,e^{i\phi_{12}}
\nonumber\\
&\quad
 - \theta_{13}\,c_{23}^\nu c_{13}^\nu\,e^{i\phi_{13}},
\label{eq:Ue3-app}\\[6pt]
U_{\mu 1} &= U^\nu_{\mu 1}
 + \theta_{12}\,e^{-i\phi_{12}}\,U^\nu_{e1}
 - \theta_{23}\,e^{i\phi_{23}}\,U^\nu_{\tau 1},
\label{eq:Um1}\\[4pt]
U_{\mu 2} &= U^\nu_{\mu 2}
 + \theta_{12}\,e^{-i\phi_{12}}\,U^\nu_{e2}
 - \theta_{23}\,e^{i\phi_{23}}\,U^\nu_{\tau 2},
\label{eq:Um2}\\[4pt]
U_{\mu 3} &= U^\nu_{\mu 3}
 + \theta_{12}\,e^{-i\phi_{12}}\,U^\nu_{e3}
 - \theta_{23}\,e^{i\phi_{23}}\,U^\nu_{\tau 3}
\nonumber\\
&= s_{23}^\nu c_{13}^\nu
 + \theta_{12}\,s_{13}^\nu\,e^{-i(\phi_{12}+\delta_\nu)}
\nonumber\\
&\quad
 - \theta_{23}\,c_{23}^\nu c_{13}^\nu\,e^{i\phi_{23}},
\label{eq:Um3}\\[6pt]
U_{\tau 1} &= U^\nu_{\tau 1}
 + \theta_{13}\,e^{-i\phi_{13}}\,U^\nu_{e1}
 + \theta_{23}\,e^{-i\phi_{23}}\,U^\nu_{\mu 1},
\label{eq:Ut1}\\[4pt]
U_{\tau 2} &= U^\nu_{\tau 2}
 + \theta_{13}\,e^{-i\phi_{13}}\,U^\nu_{e2}
 + \theta_{23}\,e^{-i\phi_{23}}\,U^\nu_{\mu 2},
\label{eq:Ut2}\\[4pt]
U_{\tau 3} &= U^\nu_{\tau 3}
 + \theta_{13}\,e^{-i\phi_{13}}\,U^\nu_{e3}
 + \theta_{23}\,e^{-i\phi_{23}}\,U^\nu_{\mu 3}
\nonumber\\
&= c_{23}^\nu c_{13}^\nu
 + \theta_{13}\,s_{13}^\nu\,e^{-i(\phi_{13}+\delta_\nu)}
\nonumber\\
&\quad
 + \theta_{23}\,s_{23}^\nu c_{13}^\nu\,e^{-i\phi_{23}}.
\label{eq:Ut3}
\end{align}
In Eqs.~\eqref{eq:Ue3-app}, \eqref{eq:Um3}, and \eqref{eq:Ut3}
we have substituted the explicit PDG forms of the third-column
neutrino elements ($U^\nu_{e3}=s_{13}^\nu e^{-i\delta_\nu}$,
$U^\nu_{\mu 3}=s_{23}^\nu c_{13}^\nu$,
$U^\nu_{\tau 3}=c_{23}^\nu c_{13}^\nu$).

\subsection{PDG observables to leading order}

The standard PDG extraction gives, to leading order in
$\theta_{ij}^e$:
\begin{align}
\sin^2\theta_{13} &= |U_{e3}|^2,
\label{eq:s213-app}\\[4pt]
\sin^2\theta_{12} &= \frac{|U_{e2}|^2}{1-|U_{e3}|^2},
\label{eq:s212-app}\\[4pt]
\sin^2\theta_{23} &= \frac{|U_{\mu 3}|^2}{1-|U_{e3}|^2}
\nonumber\\
&\simeq s_{23}^{\nu\,2}
- 2\,s_{23}^\nu c_{23}^\nu\,\theta_{23}\cos\phi_{23},
\label{eq:s223-app}
\end{align}
where in the last line we used Eq.~\eqref{eq:Um3}
and $c_{13}^\nu\simeq 1$ since $\theta_{13}^\nu\sim\e$.
Equation~\eqref{eq:s223-app} is equivalent to the atmospheric-angle
correction Eq.~\eqref{eq:theta23-theorem} in the main text.

The Dirac CP phase is extracted from the Jarlskog invariant,
\begin{equation}
\JCP = \mathrm{Im}\!\left[
U_{e1}\,U_{\mu 2}\,U_{e2}^*\,U_{\mu 1}^*
\right],
\label{eq:JCP-app}
\end{equation}
together with the relation
$\JCP = c_{12}s_{12}\,c_{23}s_{23}\,c_{13}^2s_{13}\sin\delta$.
The reactor element Eq.~\eqref{eq:Ue3-app} directly yields the
analytic $\delta$-theorem of Eq.~\eqref{eq:delta-theorem} upon
writing $U_{e3}=s_{13}\,e^{-i\delta}$.

\subsection{Summary of model-parameter counting}

The full PMNS matrix depends on 10 continuous parameters:
4 neutrino ($\theta_{12}^\nu,\theta_{23}^\nu,\theta_{13}^\nu,\delta_\nu$)
and 6 charged-lepton ($c_{12}^e,c_{23}^e,c_{13}^e,\phi_{12}^e,\phi_{23}^e,\phi_{13}^e$).
Of these, the $B$-lattice and the reflection symmetry fix:
\begin{itemize}
\item $\theta_{23}^\nu=45^\circ$ and $\delta_\nu=270^\circ$ (exact, up to the
discrete sign), and
$\theta_{12}^\nu\simeq\arctan(1/\!\sqrt{2})\simeq 35.3^\circ$
(reflection / TBM structure);
\item $\theta_{13}^\nu\sim\e$ (one free $\mathcal{O}(1)$ parameter);
\item the power-law scalings in Eq.~\eqref{eq:theta-e-scaling}
(three free $\mathcal{O}(1)$ coefficients);
\item $\phi_{12}^e\approx\phi_{23}^e$ with misalignment
$\sim\!\phi_0/9$ (single-flavon phase alignment; exact in the realization
of Appendix~\ref{app:charges-scan}).
\end{itemize}
In the scan of Sec.~\ref{sec:scan_corr}, $\delta_\nu=270^\circ$ with the
reflection-breaking band of Sec.~\ref{sec:mutau}; $\phi_{13}^e$ is
the one unconstrained phase, $\phi_{12}^e$ and $\phi_{23}^e$ are
drawn from a narrow Gaussian about a common uniform mean, and the three
$\mathcal{O}(1)$ coefficients are varied uniformly in $[0.5,\,1.5]$.

\section{Relation to the companion-paper neutrino texture}
\label{app:companion}
The companion paper~\cite{Barger2025bfnb} derives a neutrino exponent
matrix from the strict Weinberg-operator additive rule
$p^\nu_{ij}=Q(L_i)+Q(L_j)$ with lepton-doublet charges
$Q(L_i)=(1,\tfrac12,0)$, giving
\begin{equation}
p^\nu_{\text{comp.}}=
\begin{pmatrix}
2 & \tfrac32 & 1\\[2pt]
\tfrac32 & 1 & \tfrac12\\[2pt]
1 & \tfrac12 & 0
\end{pmatrix}.
\label{eq:pnu-companion}
\end{equation}
This hierarchical texture reproduces the correct mass eigenvalue
scaling $m_1:m_2:m_3\sim\e^2:\e:1$, which is the only neutrino
observable the companion paper addresses.
The present paper adopts instead the flat texture of
Eq.~\eqref{eq:pnu-texture}, in which
$p^\nu_{L_2}=p^\nu_{L_3}=0$, so that the entire $23$ block is
$\mathcal{O}(m_0)$.  This flattening is necessary to accommodate
the large atmospheric and solar mixing angles through the mu--tau
reflection symmetry of the $\mathcal{O}(1)$ coefficients.
The mass eigenvalue hierarchy is preserved: it arises from the
$\mathcal{O}(\e)$ suppression of the first-generation couplings
rather than from a hierarchical $23$ block.

The two assignments share common features:
(i)~the same expansion parameter $\e=1/B$;
(ii)~the same first-generation suppression
$p^\nu_{1j}\geq 1$;
(iii)~identical mass-eigenvalue ratios
$m_1:m_2:m_3\sim\e^2:\e:1$; and
(iv)~identical charged-lepton mass scaling
$m_e:m_\mu:m_\tau\propto\e^{29/6}:\e^{5/3}:1$.
They differ only in the $23$ sub-block of the neutrino
texture, the sector where the reflection-symmetry
input of the present paper acts.
In this sense the two treatments are complementary: the companion
paper establishes the $B$-lattice mass framework, while the present
paper introduces the symmetry structure required for the PMNS
mixing pattern.

In earlier two-over-two constructions the Majorana nature of the
neutrino operator required a charge normalization denominator of 18,
reflecting the bilinear structure of the effective dimension-five
operator.  In the present effective-texture formulation this doubling
is absorbed into the definition of $p^\nu_{ij}$, allowing a minimal
denominator of 9 without altering the predicted $B$-scaling relations
for masses or mixing.

\section{Horizontal charges, matter content, and the coefficient scan}
\label{app:charges-scan}

This appendix consolidates the inputs of the construction: the assumed
matter content, the horizontal-charge origin of the lepton textures, the
status of the $\mu$--$\tau$ symmetry, and the measure underlying the
coefficient scan.  The charge assignment is that of the published
two-over-two analysis~\cite{Barger2025bfnb}; the parts needed here are
restated so that the present paper stands on its own.

\subsection{Matter content}

The field content is that of the Standard Model with a single
Froggatt--Nielsen flavon $\Phi$ and heavy vectorlike messengers; no
light states beyond the three active neutrinos are introduced.
Light-neutrino masses come from the dimension-five Weinberg
operator~\cite{Weinberg}
\begin{equation}
\frac{1}{\Lambda_L}\,(L_i H)(L_j H)
\left(\frac{\langle\Phi\rangle}{\Lambda}\right)^{p^\nu_{ij}},
\end{equation}
so that $(\mnu)_{ij}\sim c^\nu_{ij}\,\e^{\,p^\nu_{ij}}\,m_0$ with
$m_0\sim\langle H\rangle^2/\Lambda_L$.  A high-scale completion, for
example type-I seesaw exchange of heavy gauge-singlet
neutrinos~\cite{Minkowski,Yanagida,GellMannRamond,MohapatraSenjanovic},
generates this operator but is not required; every result below uses
only the operator and its $B$-power counting and is therefore
independent of the ultraviolet origin of $\Lambda_L$.

\subsection{Horizontal charges and the lepton textures}

A single $U(1)$ horizontal symmetry assigns the lepton charges (in units
of the flavon charge)
\begin{equation}
Q(L_i)=\left(1,\tfrac12,0\right),\qquad
Q(e^c_j)=\left(\tfrac{23}{6},\tfrac76,0\right),
\label{eq:lepton-charges}
\end{equation}
The diagonal charged-lepton
exponents follow by additivity,
\begin{equation}
p^e_{ii}=Q(L_i)+Q(e^c_i)=\left(\tfrac{29}{6},\tfrac53,0\right),
\end{equation}
reproducing the mass hierarchy of Eq.~\eqref{eq:me-hierarchy}; the same
charges give the neutrino Weinberg texture $p^\nu_{ij}=Q(L_i)+Q(L_j)$ of
Eq.~\eqref{eq:pnu-companion} (Appendix~\ref{app:companion}).  The masses
and the neutrino texture are thus fixed by Eq.~\eqref{eq:lepton-charges}
alone.

Only the left-handed rotations enter $\UPMNS$, and these are set by the
column differences $p^{eL}_{ij}\equiv p^e_{ij}-p^e_{jj}$ (since
$(\Ue)_{ij}\sim\e^{\,p^{eL}_{ij}}$); the right-handed completion of $p^e$
is unobservable, and the symmetric matrix of Eq.~\eqref{eq:pe-texture} is
one convenient representative.  The observable content of $\Ue$ is
therefore the three left-handed angles
\begin{equation}
\theta^e_{12}\sim\e^{17/9},\qquad
\theta^e_{23}\sim\e^{2},\qquad
\theta^e_{13}\sim\e^{20/9}.
\end{equation}
These sit below the values that strict additivity with
Eq.~\eqref{eq:lepton-charges} would give
($\theta^e_{12},\theta^e_{23}\sim\e^{1/2}$ and $\theta^e_{13}\sim\e$); the
additional suppression of the $(1,2)$, $(1,3)$, and $(2,3)$ entries, by
$\e^{25/18}$, $\e^{22/18}$, and $\e^{27/18}$, is the assumption of
charged-lepton diagonal dominance, a set of wrinkles on the Standard-Model
Yukawa chains in the sense of Ref.~\cite{AsadiWrinkles}.  The $(2,3)$ entry
emphasized in the main text is the strongest case, where the left-handed
exponent is raised from $\tfrac12$ to $2$; this keeps the charged-lepton
correction to atmospheric mixing perturbative.  It is not a property of the
messenger content.  In a single-flavon completion with generic couplings,
every chain that generates the $(2,2)$ entry contains a chain for the
$(2,3)$ entry as a sub-chain (a chain from $L_2$ to $e^c_2$ passes the sites
at which a chain to $e^c_3$ exits, and those exit vertices are allowed), so
no pattern of absent or heavy messengers raises the $(2,3)$ exponent without
raising the $(2,2)$ exponent by at least as much, and likewise for $(1,2)$
and $(1,3)$ relative to $(1,1)$.  The suppression requires a selection rule
that distinguishes the lepton doublets by generation, i.e., an additional
symmetry under which the diagonal entries $L_iHe^c_i$ are neutral and the
off-diagonal entries are not, broken by a spurion other than the flavon.
The minimal example is a $Z_3'$ with $q'(L_i)=(0,1,2)$,
$q'(e^c_j)=(0,2,1)$, $H$ and $\Phi$ neutral, and a spurion $\eta$ of charge
$1$ with $|\langle\eta\rangle|/\Lambda=\e^{w}$; every diagonal entry is
neutral and every off-diagonal entry carries exactly one insertion of $\eta$
or $\eta^\dagger$, so the masses are unchanged, the left-handed exponents
become $(\tfrac12+w,\tfrac12+w,1+w)$, and the rotation phases are
$\phi^e_{12}=\phi^e_{23}=\arg\langle\eta\rangle=-\phi^e_{13}$ exactly.  At
the messenger level, the tower fed by $L_2$ carries the $Z_3'$ charge of
$L_2$ and cannot exit onto $e^c_3$, since the vertex $D_0He^c_3$ would carry
charge $2$; the $(2,3)$ entry arises through one heavy vectorlike pair with
the charges of $L_3$, which mixes with that tower through $\eta$ and couples
to $He^c_3$, and this pair does not enter the $(2,2)$ chain.  The
generalized CP that conjugates the $Z_3'$ charges exchanges $L_2$ and $L_3$,
i.e.\ it is the mu--tau reflection of Sec.~\ref{sec:mutau}, so the two
symmetries combine consistently.  The lattice value $w=\tfrac32$ gives
exponents $(2,2,\tfrac52)$, within the coefficient range of the scan of the
texture of Eq.~\eqref{eq:theta-e-scaling} (angle ratios $0.83$, $1$,
$0.63$); the scan with these exponents gives $\theta_{23}=43.6\pm0.7^\circ$,
$\delta=262\pm5^\circ$ and $46.3\pm0.8^\circ$, $272\pm7^\circ$ for the two
branches, against Table~\ref{tab:branch-averages}, and the finer lattice
assignment is an input at the level of the $\mathcal{O}(1)$ coefficients.
This assumption, beyond the charge set of Eq.~\eqref{eq:lepton-charges}, is
the only departure from minimal additivity in the charged-lepton sector.
Because the symmetry acts on the doublets it also acts on the operator that
generates the neutrino masses, where the insertions it requires are
reflection symmetric but would suppress the diagonal $23$ entries if small;
the $\mathcal{O}(1)$ block of the flat texture is taken in this paper from
the reflection symmetry (Appendix~\ref{app:companion}) rather than from a
chain counting, and a joint completion of the two sectors is not attempted
here.

\subsection{Status of the \texorpdfstring{$\mu$--$\tau$}{mu-tau} reflection symmetry}

Mu--tau reflection~\cite{HarrisonScott,GrimusLavoura,MuTauReview}, the
exchange $\nu_\mu\leftrightarrow\nu_\tau$ combined with CP, supplies the
$\mathcal{O}(1)$ structure of the neutrino mass matrix that the $B$-power
counting leaves free [Eq.~\eqref{eq:mutau-coeff}].  Four points fix its
status.  First, it is distinct from, and additional to, the horizontal
$U(1)$; it is not the flavon selection rule and does not enter the charge
counting of Eq.~\eqref{eq:lepton-charges}.  Second, it is a generalized CP
transformation, a discrete exchange combined with complex conjugation, not
the continuous combination $U(1)_{L_\mu-L_\tau}$, and it is not gauged; in
the residual-symmetry language of Refs.~\cite{FHZ,HLS} it is
$G_\nu=Z_2\times{\rm CP}$.  Third, it is exact at leading order in the
neutrino sector, where it fixes $\theta^\nu_{23}=45^\circ$ and
$\delta_\nu=\pm90^\circ$ and leaves $\theta^\nu_{13}$ free; its real
limit, mu--tau exchange, would force $\theta_{13}=0$.  It is broken in the
PMNS matrix by the charged-lepton rotations, which produce the octant and
Dirac-phase displacements, and in the neutrino sector by the real parts of
the coefficient splittings and by a departure of $\phi_0$ from $\pi$
(Sec.~\ref{sec:mutau}), which the scan covers with the band
$\delta_\nu=270^\circ\pm16^\circ$.  Fourth, the flavor group $S_4$
combined with generalized CP realizes it and in addition pulls the solar
angle toward its tribimaximal value~\cite{FHZ}; the present paper uses only
the residual symmetry and not a specific flavor group.

\subsection{The coefficient scan and what it fixes}

The scan samples the $\mathcal{O}(1)$ data left free once the $B$-lattice
and the reflection-symmetry input are imposed; it is a sampling of the
residual coefficients of the model, not a fit to data.  Held fixed are the
neutrino angles, $\theta_{12}^\nu=34^\circ$ near the NuFIT~6.0~\cite{NuFIT60}
normal-ordering best fit, $\theta_{23}^\nu=45^\circ$ (exact, from
reflection), and $\theta^\nu_{13}=0.158$, together with the $B$-lattice
power scalings of the three charged-lepton angles
[Eq.~\eqref{eq:theta-e-scaling}].  The neutrino phase is
$\delta_\nu=270^\circ$ from reflection; the breaking effects of
Sec.~\ref{sec:mutau} are covered by a second scan with
$\delta_\nu=270^\circ+\kappa$, $\kappa$ uniform on $[-16^\circ,16^\circ]$.
Varied are the three coefficients $c^e_{ij}$, drawn uniformly on $[0.5,1.5]$,
together with the phases.  The charged-lepton phases are drawn as
$\phi^e_{12}=\mu+g_{12}$ and $\phi^e_{23}=\mu+g_{23}$, where $g_{12}$ and
$g_{23}$ are independent Gaussian variates of zero mean and width
$\sigma=20^\circ$, the single-flavon misalignment $\phi_0/9$ at
$\phi_0\sim\pi$ derived in Sec.~\ref{sec:phase-alignment}, and the common
mean $\mu$ is uniform on $[0,2\pi)$; no central value is preferred.  The
phase $\phi^e_{13}$ is uniform on $[0,2\pi)$.  The NuFIT~6.0 $2\sigma$ cuts on
$\sin^2\theta_{12}$ and $\sin^2\theta_{13}$ are then applied, leaving the
points tabulated in Table~\ref{tab:branch-averages}.
Because the octant is the sign of $\cos\phi^e_{23}$
[Eq.~\eqref{eq:theta23-theorem}], the uniform mean populates both octants
equally before cuts; the preference for the lower octant in
Table~\ref{tab:branch-averages} is produced by the cuts on
$\sin^2\theta_{12}$ and $\sin^2\theta_{13}$, which act on the phases through
the interference in $U_{e2}$ and $U_{e3}$, and depends on the measure.
With $\delta_\nu$ fixed, the residual dispersion of $\delta$
($\sigma_\delta\simeq7$--$10^\circ$) is set by the
charged-lepton interference of Eq.~\eqref{eq:delta-theorem}: the retained
$\theta_{12}^e$ term, with its phase $\Phi_e$ scanned through the
$\mathcal{O}(1)$ coefficient spread, gives $\sigma_\delta\simeq r/\sqrt{2}\approx8^\circ$
at $c^e_{12}=1$; the subleading
$\theta_{13}^e c_{23}^\nu$ correction with its uniform phase $\phi^e_{13}$
adds a further spread, and the full range of the scan is
$240^\circ$--$296^\circ$ ($226^\circ$--$308^\circ$ with the breaking band).

Two statements of different status follow.  The relative population of the
two octants favors the lower octant but depends on the
measure, i.e.\ on the flat coefficient prior and the alignment width; it
should be read as a mild prior, not a sharp prediction, and a different
prior would shift the ratio.  The coefficient range also sets the width of
the atmospheric band: with $c^e_{23}\in[0.5,1.5]$ the scan spans
$41.7^\circ\le\theta_{23}\le48.3^\circ$ ($0.44\le\sin^2\theta_{23}\le0.56$);
widening the range to $[0.5,2]$ extends it to about $41^\circ$--$49^\circ$
($0.43$--$0.58$) without changing the branch medians, which are set by the
typical coefficient, not its extreme.  The NuFIT~6.0 upper-octant minimum,
$\sin^2\theta_{23}\approx0.56$, lies at the boundary of the adopted range.
What does not depend on the measure is the
correlation.  The octant shift [Eq.~\eqref{eq:theta23-theorem}] and the
Dirac-phase shift [Eq.~\eqref{eq:delta-theorem}] both descend from the
single interference parameter that fixes $U_{e3}$, so the sign of
$(\theta_{23}-45^\circ)$ and the direction of $(\delta-\delta_\nu)$ are
correlated.  Changing the priors moves and reweights points along the two
branches but does not remove the branches or alter the octant--$\delta$
linkage.  That linkage, and not the population ratio, is the prediction
tested by a joint determination of $\theta_{23}$ and $\delta$.

Table~\ref{tab:phase-variants} compares the default phase prescription with
two in which the charged-lepton phases are tied to the flavon phase,
$(\phi^e_{12},\phi^e_{23},\phi^e_{13})=(\tfrac{17}{9},2,\tfrac{20}{9})\phi_0$
with $\phi_0$ uniform on $[0,2\pi)$, with and without a $20^\circ$ Gaussian
smearing of each phase.  Tying the phases sharpens the branches, makes the
sign relation of Eq.~\eqref{eq:octant-quad} exact, and lowers the population
ratio to $N_{\rm LO}/N_{\rm HO}\simeq1.1$.  In the strict case the cuts accept
$\phi_0$ only within about $\pm15^\circ$ of $0^\circ$ and $180^\circ$
(lower octant) and of $90^\circ$ and $270^\circ$ (upper octant), the four
points at which the interference phase $\Phi_e=\tfrac{17}{9}\phi_0-270^\circ$
returns $|U_{e3}|$ to its measured value; the consequences for the neutrino
sector are discussed in Sec.~\ref{sec:phase-alignment}.

\begin{table}[htbp]
\centering
\caption{Dependence of the scan on the charged-lepton phase prescription, at
$\delta_\nu=270^\circ$ with the $2\sigma$ cuts of Table~\ref{tab:branch-averages}.
``Default'' is the prescription of Table~\ref{tab:branch-averages}
($\phi^e_{12}$, $\phi^e_{23}$ Gaussian about a common uniform mean, $\phi^e_{13}$ uniform);
``tied'' sets $(\phi^e_{12},\phi^e_{23},\phi^e_{13})=(\tfrac{17}{9},2,\tfrac{20}{9})\phi_0$
with $\phi_0$ uniform and a $20^\circ$ Gaussian smearing of each phase; ``strict'' uses the
same relations with no smearing, so that the phases are functions of $\phi_0$ alone.
Entries are the branch medians of $\theta_{23}$, the branch means and standard deviations of
$\delta$ (degrees), the population ratio, and the fraction of points obeying the sign relation of
Eq.~\eqref{eq:octant-quad}. In the strict case the cuts accept $\phi_0$ only within about
$\pm15^\circ$ of $0^\circ$ and $180^\circ$ (lower octant) and of $90^\circ$ and $270^\circ$
(upper octant).  The strict variant is the limit in which every charged-lepton
phase traces to the flavon phase $\phi_0$; the realization of
Sec.~\ref{app:charges-scan} gives instead $\phi^e_{12}=\phi^e_{23}=-\phi^e_{13}$
with a free common phase, which the default prescription represents.}
\label{tab:phase-variants}
\setlength{\tabcolsep}{4pt}
\renewcommand{\arraystretch}{1.1}
\resizebox{\columnwidth}{!}{%
\begin{tabular}{lcccc}
\hline\hline
Prescription & $\theta_{23}$ (LO/HO) & $\delta$ (LO/HO) & $N_{\rm LO}/N_{\rm HO}$ & sign rel.\ [Eq.~\eqref{eq:octant-quad}] \\
\hline
default & $43.7^\circ/46.2^\circ$ & $260\pm7/274\pm10$ & $1.34$ & $0.80$ \\
tied, $20^\circ$ smearing & $43.2^\circ/46.8^\circ$ & $256\pm6/280\pm5$ & $1.08$ & $0.98$ \\
strict, $\phi_0$ only & $43.0^\circ/47.0^\circ$ & $255\pm5/281\pm4$ & $1.12$ & $1.00$ \\
\hline\hline
\end{tabular}}
\end{table}

\section{Absence of dangerous charged-lepton flavor violation}
\label{app:fcnc}

The charged-lepton rotations $U_{eL}$ and $U_{eR}$ are not an independent
source of flavor violation.  With a single Higgs doublet the charged-lepton
Yukawa matrix and the mass matrix are proportional, so the rotation that
diagonalizes the masses also diagonalizes the Higgs coupling; the right-handed
rotation $U_{eR}$ is unobservable, and $U_{eL}$ enters only through the PMNS
combination $U_{\rm PMNS}=U_e^\dagger U_\nu$ [Eq.~\eqref{eq:PMNS-factorize}].
There is accordingly no tree-level charged-lepton flavor-changing neutral
current.  The decay $\mu\to e\gamma$ proceeds only through the GIM-suppressed
light-neutrino loop, with a branching ratio of order $10^{-54}$, far below the
present MEG~II limit $\mathcal{B}(\mu^+\to e^+\gamma)<1.5\times10^{-13}$ at
$90\%$ C.L.~\cite{MEGII2025}.

Higher-dimension operators generated by integrating out the flavon and the
messengers at the scale $\Lambda$ (with $\langle\Phi\rangle/\Lambda=\e$) include
a dipole term $(m_\mu/\Lambda^2)\,\bar e_L\,\sigma^{\mu\nu}\mu_R\,F_{\mu\nu}$
that mediates $\mu\to e\gamma$; it is suppressed by $1/\Lambda^2$ and by the
chirality factor $m_\mu$, and the same $1/\Lambda^2$ suppression governs the
tree-level exchange of the heavy messengers.  For $\Lambda$ well above the
electroweak scale the induced $\mu\to e\gamma$ rate, and the related
$\mu\to 3e$ and $\mu$--$e$ conversion rates, lie far below their experimental
bounds.  Realized as the discrete $Z_N$ of Sec.~\ref{sec:discrete}, rather than
a spontaneously broken continuous global symmetry, the horizontal symmetry
produces no Goldstone familon, so the bounds on $\mu\to e+\text{familon}$ do not
apply.  The one assumption underlying these statements is
$\Lambda\gg\text{TeV}$, standard for Froggatt--Nielsen constructions; a
low-scale completion, a second Higgs doublet, or supersymmetric sleptons would
introduce additional flavor-violating couplings and would call for a separate
analysis.

\begin{acknowledgments}
V.B. gratefully acknowledges support from the U.S. Department of Energy, Office of Science, Office of High Energy Physics, under Award Number DE-SC0017647 and from the William F. Vilas Estate.
\end{acknowledgments}

\end{document}